# $\gamma$-phase Inclusions as Common Defects in
# Alloyed $\beta$-(Al$_x$Ga$_{1−x}$)$_2$O$_3$ and Doped $\beta$-Ga$_2$O$_3$ Films

Celesta S. Chang[1,2,†], Nicholas Tanen[3,†], Vladimir Protasenko[4], Thaddeus J. Asel[5],

Shin Mou[5], Huili Grace Xing[3,4,6], Debdeep Jena[3,4,6], and David A. Muller[2,6,*]

[1]*Department of Physics, Cornell University, Ithaca, NY 14853, USA*

[2]*School of Applied and Engineering Physics, Cornell University, Ithaca, NY 14853, USA*

[3]*Department of Materials Science and Engineering, Cornell University, Ithaca, NY 14853, USA*

[4]*School of Electrical and Computer Engineering, Cornell University, Ithaca, NY 14853, USA*

[5]*Air Force Research Laboratory, Materials and Manufacturing Directorate, Wright Patterson AFB, Ohio 45433, USA*

[6]*Kavli Institute at Cornell for Nanoscale Science, Cornell University, Ithaca, NY 14853, USA*

[†] Equal Contribution

[*]Corresponding author: dm24@cornell.edu

## ABSTRACT

$\beta$-Ga$_2$O$_3$ is a promising ultra-wide bandgap semiconductor whose properties can be further enhanced by alloying with Al. Here, using atomic-resolution scanning transmission electron microscopy (STEM), we find the thermodynamically-unstable $\gamma$-phase is a ubiquitous defect in both $\beta$-(Al$_x$Ga$_{1−x}$)$_2$O$_3$ films and doped $\beta$-Ga$_2$O$_3$ films grown by molecular beam epitaxy. For undoped $\beta$-(Al$_x$Ga$_{1−x}$)$_2$O$_3$ films we observe $\gamma$-phase inclusions between nucleating islands of the $\beta$-phase at lower growth temperatures (~400-600 °C). In doped $\beta$-Ga$_2$O$_3$, a thin layer of the $\gamma$-phase is observed on the surfaces of films grown with a wide range of n-type dopants and dopant concentrations. The thickness of the $\gamma$-phase layer was most strongly correlated with the growth temperature, peaking at about 600 °C. Ga interstitials are observed in $\beta$-phase, especially near the interface with the $\gamma$-phase. By imaging the same region of the surface of a Sn-doped $\beta$-(Al$_x$Ga$_{1−x}$)$_2$O$_3$ after ex-situ heating up to 400 °C, a $\gamma$-phase region is observed to grow above the initial surface, accompanied by a decrease in Ga interstitials in the $\beta$-phase. This suggests that the diffusion of Ga interstitials towards the surface is likely the mechanism for growth of the surface $\gamma$-phase, and more generally that the more-open $\gamma$-phase may offer diffusion pathways to be a kinetically-favored and early-forming phase in the growth of Ga$_2$O$_3$.

# INTRODUCTION

$Ga_2O_3$ is known to have at least five different phases, with formation energies ordered as $\beta$(C2/m) < $\kappa$(Pna2$_1$) < $\alpha$(R$\bar{3}$c) < $\delta$(Ia$\bar{3}$) < $\gamma$(Fd$\bar{3}$m) at temperatures below 1600 K [1,2]. Among these, $\beta$-$Ga_2O_3$ has received primary focus in the power electronics community for its thermodynamic stability [1-3]. Due to its ultra-wide band gap (4.4-4.9 eV), availability of high-quality melt-grown substrates including semi-insulating to heavily n-type doped wafers, early demonstrations of both lateral and vertical devices with breakdown voltages as high as 3 kV were achieved [4-11].

On the other hand, being the least stable phase in the $Ga_2O_3$ family, far less is known about $\gamma$-$Ga_2O_3$ and studies have been relatively limited. Owing to its mesoporous crystal structure, $\gamma$-$Ga_2O_3$ has been studied as a catalyst when grown in nanorods [12]. Epitaxial thin films have been grown on substrates such as MgO or sapphire, where dopants are known to provide phase stability [5,13,14]. However, $\gamma$-$Ga_2O_3$ has unique features, such as a cubic defective spinel structure with two cation vacancies in the space group Fd$\bar{3}$m [15, 16]. This can provide a platform for developing new functionalities that can arise by heteroepitaxial growth on spinel oxides. The direct and indirect energy bandgaps of $\gamma$-$Ga_2O_3$ are $E_g$=5.0 eV and 4.4 eV respectively [14], slightly higher than that of $\beta$-$Ga_2O_3$ ($E_g$=4.4-4.9 eV). Bandgap engineering by growth of $\gamma$-$(Al_xGa_{1-x})_2O_3$ has been reported to achieve gaps up to 6.97 eV by Oshima et. al.[17], suggesting its potential as a future generation ultra-wide bandgap semiconductor.

Unlike traditional semiconductor materials such as Si, Ge, and diamond (C) [18], wurtzite (III-N) [19] and zincblende (GaAs) [20] structures, which are known to be significantly more stable thermodynamically than their polymorphs, metastable phases of $Ga_2O_3$ are relatively close to $\beta$-$Ga_2O_3$ in the phase diagram. As a result, many studies have reported coexistence of two or more phases arising during growth of thin film $Ga_2O_3$ [21-25]. A recent study also reported the possible formation of $\gamma$-$(Al_xGa_{1-x})_2O_3$ in metal organic chemical vapor deposition (MOCVD) grown $\beta$-$(Al_xGa_{1-x})_2O_3$ at Al concentrations of 27%<x<40% [25], although an alternate interpretation of their TEM image showing $\gamma$-$(Al_xGa_{1-x})_2O_3$ in terms of stacking faults in the $\beta$-phase has also been made [26-28]. In this study, we observe $\gamma$-$Ga_2O_3$ inclusions as a common defect that is broadly found in series of doped $\beta$-$Ga_2O_3$ and $\beta$-$(Al_xGa_{1-x})_2O_3$, at compositions of interest for improved device performance and higher breakdown voltages.

Because $\beta$-$Ga_2O_3$ is used for power electronic applications, the observation of the $\gamma$-phase poses an interesting question as to its potential impact on the progress in this field. $\gamma$-phase inclusions in $\beta$-phase structure could be detrimental for device performance. $\gamma$-phase at the surface of doped films can lead to non-ideal high current ohmic contact regions, while $\gamma$-$(Al_xGa_{1-x})_2O_3$ in $\beta$-$(Al_xGa_{1-x})_2O_3$ could severely compromise the gate switching behaviors under high-field. Additionally, as large voltages and currents used in power electronics can lead to significant joule heating of devices without proper thermal management, device reliability can be harmed by $Ga_2O_3$ polymorphs that could be transformed at elevated temperatures. More understanding of $\gamma$-$Ga_2O_3$ is needed to measure the extent to which it can affect the device performance and to prevent its unintentional formation in $\beta$-$Ga_2O_3$.

In this paper, we use molecular beam epitaxy (MBE) to grow crystalline thin films of Ge-, Si-doped $\beta$-$Ga_2O_3$, Sn-doped $\beta$-$(Al_{0.15}Ga_{0.85})_2O_3$, and undoped series of $\beta$-$(Al_xGa_{1-x})_2O_3$. Since MBE growth occurs in ultrahigh vacuum systems using high purity source materials, it allows for high purity, highly crystalline thin films with atomically smooth and abrupt interfaces which are necessary for high performance electronic devices. We then characterize the films using atomic force microscopy (AFM), high resolution x-ray diffraction (HRXRD), and scanning transmission electron microscopy (STEM). Here, we find that $\gamma$-$Ga_2O_3$ is present at the surfaces of doped $\beta$-$Ga_2O_3$, while $\gamma$-$(Al_xGa_{1-x})_2O_3$ in $\beta$-$(Al_xGa_{1-x})_2O_3$ is also found as inclusions inside the film at low growth temperatures. Expanding the phase space studied by [25], here we report that even at relatively low Al concentrations ($< 20\%$), the inclusion of $\gamma$-$(Al_xGa_{1-x})_2O_3$ can be observed, and therefore it appears that the formation of $\gamma$-$(Al_xGa_{1-x})_2O_3$ is also sensitive to growth conditions rather than solely to the Al composition. We suggest that the $\gamma$-phase inclusions, especially near the surface, may be stabilized by the presence of dopants during growth. In the last section, we focus on the Ga-interstitials that are frequently observed near the formation of $\gamma$-phase in both alloyed and doped $\beta$-phase films. From our ex-situ heating results, we suggest the possibility of Ga interstitials participating in the formation of $\gamma$-phase by out diffusion, which suggests a mechanism for why the thermodynamically-unstable $\gamma$-phase could be a kinetically-preferred phase, especially for lower-temperature growth of $Ga_2O_3$.

## METHODS

Two Ge-doped $\beta$-$Ga_2O_3$ thin films were grown using a Veeco GEN Xcel plasma assisted MBE (PAMBE) system in air force research laboratory (AFRL), equipped with standard effusion cells for Ge and Ga and an unibulb RF plasma source for oxygen. The films were grown with substrate temperatures of 550 °C and 650 °C, an oxygen flow rate of 1.55 sccm which resulted in a background pressure of $1.5 \times 10^{-5}$ torr and 300 W of oxygen plasma power. Ge was maintained at 600 °C for all growths and the Ga beam equivalent pressure was $6.0 \times 10^{-8}$ torr which is in the O-rich regime for our system.

Ge-doped $\beta$-Ga$_2$O$_3$ (600 °C), Sn-doped $\beta$-(Al$_{0.15}$Ga$_{0.85}$)$_2$O$_3$, Si-doped $\beta$-Ga$_2$O$_3$, and series of $\beta$-(Al$_x$Ga$_{1-x}$)$_2$O$_3$ films were grown using a Veeco Gen930 MBE system at Cornell University. The activated oxygen flux was provided using a Veeco RF plasma source. Al and Ga metals were provided using standard Veeco effusion cells. The films were all grown on edge-defined film-fed grown (EFG) (010) substrates purchased from Novel Crystal Technology (NCT) which were diced into 5 mm × 5 mm pieces and In-mounted on a two-inch silicon carrier wafer. Prior to growth, the substrates were cleaned using a degreasing soap and deionized water followed by sonication in acetone, methanol, isopropyl alcohol and again in deionized water. The films were annealed for 30 minutes at 800 °C (as measured by a thermocouple) while being exposed to an oxygen plasma prior to growth to descum the growth surface. The Si- and Ge-doped (600 °C) samples were prepared in a similar way; however, no plasma descum was performed for these two samples. For the Sn-doped sample, purchased from NCT, effusion cells were used to provide Ga, Al and SnO$_2$. Specific growth parameters for each sample and sample labels can be found in Table I. In order to obtain precise Al and Sn concentrations, calibration samples were grown. The Al concentration was determined using Rutherford back scattering (RBS), and the Sn concentration was determined using electrochemical capacitance voltage (ECV) measurements to profile N$_d$-N$_a$ versus depth.

The $\beta$-(Al$_x$Ga$_{1-x}$)$_2$O$_3$ films in this experiment were grown at 500 °C, 600 °C and 700 °C (as measured by the MBE CAR thermocouple). The alloyed samples are also labelled as A1-A3 with ascending growth temperature. Samples A2 and A3 were grown using a Ga beam equivalent pressure (BEP) of $2 \times 10^{-8}$ Torr (flux of $5.83 \times 10^{13}$ atoms/cm$^2$s) and an Al BEP of $3.5 \times 10^{-9}$ Torr ($8.04 \sim 8.75 \times 10^{12}$ atoms/cm$^2$s) in order to provide approximately 15 % Al based on total metal fluxes. The oxygen flow rate of 0.57 sccm resulted in a chamber pressure of $1 \times 10^{-5}$ Torr and the RF forward power for the plasma supply was set to 250W. We chose the Ga flux which maximized the growth rate in our system and should correspond to approximately the stoichiometric point as described by Vogt et. al. [29].

It should be noted that these $\beta$-(Al$_x$Ga$_{1-x}$)$_2$O$_3$ films were grown at relatively low growth rates (32-55 nm/hr) for (010) $\beta$-Ga$_2$O$_3$ films grown by PAMBE. In comparison, other PAMBE studies typically report growth rates between 120-180 nm/hr [30, 31], and over 600 nm/hr was reported for ozone-MBE [32], while a growth rate of 1.6 $\mu$m/hr was recently achieved for suboxide-MBE [33]. Growth rates of almost 20 $\mu$m/hr and 0.82-1.47 $\mu$m/hr can be achieved using higher pressure deposition techniques like halide vapor phase epitaxy (HVPE) [34] and MOCVD [25], respectively. The low growth rate in our study was a consequence of our oxygen plasma source and system vacuum pumping capacity which sets an upper limit on the supplied, activated oxygen and thus the achievable growth rates.

HRXRD measurements were performed on a Panalytical X'pert Pro x-ray diffractometer to obtain film thicknesses and Al concentrations. To study surface morphology, AFM measurements were performed on a Veeco Icon system. Table 1 also summarizes the Al content extracted from XRD measurements, and AFM root mean square (RMS) roughness for the samples discussed in this paper.

Cross-sectional TEM specimens were prepared using a FEI Strata 400 Focused Ion Beam (FIB) and Helios G4-UX FIB with a final milling step of 5 keV to reduce damage. To prevent surface damage from the ion-beam, carbon and platinum protective layers were deposited prior to milling. The samples were then examined by high-angle annular dark field scanning transmission electron microscopy (HAADF-STEM) imaging, using an aberration-corrected Titan Themis operating at 300 keV.

To investigate the relevance of Ga interstitials to $\gamma$-phase formation, three Sn-doped $\beta$-$(Al_{0.15}Ga_{0.85})_2O_3$ TEM lamellae were heated twice, from room temperature up to 200 °C then 400 °C in an Argon environment using Thermocraft XST-2-0-12-1V2-F04. Ramping rate of 5 °C /min was used and the highest temperature was maintained for 2 hours. Cooling was let to happen naturally until it reached room temperature, which approximately corresponds to a cooling rate of 30 °C /min.

## RESULTS AND DISCUSSIONS

### Distinguishing the crystal Structure between $\gamma$-$Ga_2O_3$ and superimposed $\beta$-$Ga_2O_3$

Determining the crystal phase simply based on the atomic arrangement seen in STEM images can be difficult and sometimes misleading if the image resolution is too low to distinguish between similar crystal structures. In a previous study by Bhuiyan et.al. [25] they reported the presence of $\gamma$-$(Al_xGa_{1-x})_2O_3$ with x-ray diffraction, as well as STEM imaging. However, a comment by Wouters et. al. [26] argued that their STEM images instead showed a superimposed crystal structure created by stacking faults of $\beta$-$(Al_xGa_{1-x})_2O_3$ that is superficially similar to the $\gamma$-phase. The discussion on distinguishing the two different structures in the comment and the response were made based on the atomic site intensities, and image simulations which were compared with a low signal-to-noise high-angle annular dark field (HAADF)-STEM image [25-28].

Here, we show that by using an aberration-corrected microscope in a low-noise environment with a combination of imaging modes, we can obtain high-resolution, high signal-to-noise images that eliminates any ambiguity in determining the crystal phases. In HAADF-STEM, the atomic-number (Z) sensitivity roughly scales as $Z^{1.7}$, so heavy Ga atoms show up in the image while light oxygen is difficult to see [35]. On the other hand, annular bright field (ABF)-STEM imaging has a weaker dependence on the atomic number ($\sim Z^{0.8}$), thus it can easily capture oxygen sublattices in the presence of relatively heavy Ga atoms [36]. Combining these two imaging modes, we can construct a full crystal structure enabling easier determination of crystal phases. Figure 1 shows the crystal structures for $\beta$-Ga$_2$O$_3$ and $\gamma$-Ga$_2$O$_3$ in different crystallographic zones. The crystal model for $\beta$-Ga$_2$O$_3$ was constructed based on [37] while the Fe$_3$O$_4$ structure [38] was used for $\gamma$-Ga$_2$O$_3$ as they are in the same space group [39]. Different colors are used in the crystal model to distinguish the octahedral Ga sites (yellow) from tetrahedral Ga sites (blue). Simultaneously-acquired HAADF-STEM, ABF-STEM images of Sn-doped $\beta$-(Al$_{0.15}$Ga$_{0.85}$)$_2$O$_3$ and a combined false-color overlay of the two are shown for each crystal structure. The combined image is obtained by color-coding the HAADF image showing Ga atoms in green, and the ABF image showing O atoms in red by overlaying the HAADF-STEM image on an inverted ABF-STEM image.

$\beta$-Ga$_2$O$_3$ with the space group C2/m has a lattice parameter of a=12.214 Å, b=3.037 Å, c=5.798 Å [39]. $\beta$-Ga$_2$O$_3$ is generally grown parallel to the (010) surface as in Fig. 1(a), because the symmetry of the growth surface prevents twin formation, and the growth rate is much faster than that for other substrate orientations in similar MBE growth conditions [40-42]. Observation along the crystal zone of [001] in Fig. 1(b) shows the outlined hexagonal crystal structure, where Ga occupies four tetrahedral sites and two octahedral sites. In contrast, $\gamma$-Ga$_2$O$_3$ has a defective cubic spinel structure with the space group Fd$\bar{3}$m as shown in Fig. 1(c). It has a cubic lattice parameter of a=8.23 Å [15, 16], approximately three times larger in volume and 7 % lower in atomic density compared to the $\beta$-phase [39, 43]. $\gamma$-Ga$_2$O$_3$ has six tetrahedral Ga sites and five octahedral Ga sites inside a hexagonal structure overlaid in Fig. 1(c). Our observations in the next section will show that the $\gamma$-phase tends to grow on the surface of $\beta$-(010), starting from the two tetrahedral sites in the small hexagonal structure of $\beta$-Ga$_2$O$_3$. The distance between the two nearest tetrahedral sites in $\beta$-phase is 3.612 Å, while that of $\gamma$-phase is 3.564 Å. Because of this 1.3 % lattice mismatch, the growth of $\gamma$ on $\beta$ would ultimately entail the formation of domain boundaries or shifts of the $\gamma$ structure. As a TEM image captures the projection of all atoms in the column along a thickness of approximately 20-100 nm, shifts of lattice planes at different thicknesses would also result in a unique signature structure in projection. Such overlap results in Fig. 1(d), where three Ga atoms -octahedral Ga sitting in between two tetrahedral Ga sites- form a diagonal, ladder-like crystal structure, which we see often in our films.

Similar to the superimposed structure in Fig. 1 (d), a combined lattice shift of $\beta$-Ga$_2$O$_3$ [010] sheets across a stacking fault in the <102> direction [26, 27] can create new superimposed structures. These structures, as shown in Fig. 2 (b) and (c), superficially resembles that of $\gamma$-Ga$_2$O$_3$ in Fig. 2 (a) as they all have a bright atomic site in the middle. Thus, when seen in noisy or low-resolution TEM images, one can easily misinterpret the phases. The high-resolution HAADF-STEM images show a clear difference between the three crystal structures. The 2D-projected repeat unit for each crystal structure is outlined by yellow dotted lines on the HAADF-STEM images and black solid lines in the crystal model. $\gamma$-Ga$_2$O$_3$ is characterized by edge-sharing hexagonal beehive structure where the four side-edges have three Ga atoms lined up in a straight line. The outline of the superimposed structure in Fig. 2 (b) may seem to have a similar hexagonal repeat unit with the octahedral site in the middle, but they have rounded side-edges, and the actual projected repeat unit is a parallelogram. The structure in Fig. 2 (c) has four atoms in the side-edges and has a distorted hexagonal repeat unit in projection. The different locations of the oxygen atoms also reinforce these differences. Based on understanding the differences of the crystal structures, we determined $\gamma$-phase inclusions in our series of samples discussed in this paper.

## Doped $\beta$-Ga$_2$O$_3$

Three Ge-doped $\beta$-Ga$_2$O$_3$ thin films were each grown at different temperatures of 550 °C (DG1), 600 °C (DG2), and 650 °C (DG3) to understand the temperature effects on the growth and film quality (See Table I for details and film labels). The carrier concentration was on the order of $5\times10^{18}$ cm$^{-3}$ and $\sim10^{17}$ cm$^{-3}$ respectively for DG1 and DG3 grown films. DG2 was aimed for $\sim10^{18}$ cm$^{-3}$ density, however it was non-conductive and therefore the carrier concentration could not be extracted. Figure 3. shows the growth schematics for doped samples. The substrates were Fe-doped EFG $\beta$-Ga$_2$O$_3$, and as the thickness of DG2 is smaller than the other two, it was grown on a buffer layer to prevent the diffusion of Fe-dopants during growth at high temperatures [44, 45].

Figure 4. (a)-(b), (c)-(d), (e)-(f) shows HAADF-STEM images of the Ge-doped $\beta$-Ga$_2$O$_3$ films (DG1-DG3) grown at 550 °C, 600 °C, and 650 °C, respectively. Here we note that the surface layer in DG1 and DG2 films are region dependent, meaning they are not resolved everywhere on the film, but can be observed in most places (Fig. S1). High magnification HAADF-STEM images in Fig. 4 (a), (c), (e) shows all three films having $\gamma$-phase structures on the surface of the films. The thickness of the $\gamma$-phase

surface layer is 5 nm, 7 nm, and 1 nm on average for Ge-doped $\beta$-Ga$_2$O$_3$ films grown at 550 °C, 600 °C, and 650 °C, respectively. The shallow depth of the $\gamma$-phase layer in 650 °C grown film (DG3) might be explained as a higher growth temperature suppressing the formation of $\gamma$-phase while thermodynamically stabilizing the $\beta$-phase. This could also be a direct effect of the film having a lower Ge-doping concentration of ~10$^{17}$ cm$^{-3}$, however we note that for the silicon doping discussed below, the same layer thickness was seen for ~10$^{18}$ cm$^{-3}$ vs ~10$^{16}$ cm$^{-3}$ Si dopant concentration when grown at the same temperature.

To determine the role of dopants in the formation of the $\gamma$-phase, we extended our study to grow films with other dopants. Two samples, Sn-doped $\beta$-(Al$_{0.15}$Ga$_{0.85}$)$_2$O$_3$ (DSn) and Si-doped $\beta$-Ga$_2$O$_3$ (DSi) were grown on buffer layers and EFG (010) $\beta$-Ga$_2$O$_3$ substrates at a growth temperature of 550 °C and 650 °C respectively. The carrier concentration for DSn and DSi was 2×10$^{18}$ cm$^{-3}$ and 2×10$^{20}$ cm$^{-3}$ respectively. We also examined an unintentionally doped (UID) $\beta$-Ga$_2$O$_3$ films grown at 650 °C with a Si dopant concentration of ~10$^{16}$ cm$^{-3}$. In low magnification TEM image (Fig. 5 (b)) we can see the formation of a distinct surface layer on Sn-doped $\beta$-(Al$_{0.15}$Ga$_{0.85}$)$_2$O$_3$ (DSn) similar to the Ge-doped samples grown at 550 °C (DG1) and 600 °C (DG2). On the surface of Si-doped $\beta$-Ga$_2$O$_3$ (DSi) in Fig. 5 (h), we can see an onset of $\gamma$-phase formation by the appearance of the octahedral Ga site in the middle of the hexagon. The UID $\beta$-Ga$_2$O$_3$ also grown at 650 °C shows a thin $\gamma$-phase at the surface (Fig. S2). This is also similar to Ge-doped $\beta$-Ga$_2$O$_3$ also grown at 650 °C (DG3).

Films with a thick layer of $\gamma$-phase show that the $\gamma$-phase tends to nucleate from its (111) plane, which corresponds to the (21$\bar{1}$) plane of $\beta$-Ga$_2$O$_3$ as seen in Fig. 4 (c). We also add that at the interfaces of $\gamma$ and $\beta$-Ga$_2$O$_3$ in all of the doped samples, we observe high densities of interstitials in $\beta$-Ga$_2$O$_3$, which we believe is correlated to the $\gamma$-phase formation. We will discuss this in more detail in the last section.

It is important to note here that our series of undoped $\beta$-(Al$_x$Ga$_{1-x}$)$_2$O$_3$ films discussed later in this study did not show any surface layers of $\gamma$-Ga$_2$O$_3$. However, Sn-doped $\beta$-(Al$_{0.15}$Ga$_{0.85}$)$_2$O$_3$ does show $\gamma$-Ga$_2$O$_3$ on the surface. This suggests that Sn-dopants may contribute to promoting surface $\gamma$-Ga$_2$O$_3$ growth. In fact, while the previous three Ge-doped samples showed the formation of $\gamma$-phase only on the surface of the sample, Sn-doped $\beta$-(Al$_{0.15}$Ga$_{0.85}$)$_2$O$_3$, which was a low-temperature growth, also shows $\gamma$-like defects appearing inside the columns of $\beta$-(Al$_{0.15}$Ga$_{0.85}$)$_2$O$_3$. In Fig. 5 (d) this region is pointed out by a red arrow. A crystal structure from Fig. 1 (c) is overlaid as guide to the eye.

We also note that in Fig. 5 (d) there are octahedral Ga sites with brighter contrast than neighboring Ga sites, which are noted by green arrows. These bright octahedral Ga sites can be found close to the surface (Image with a wider field of view is provided as Fig. S3 in the supplementary). Sn is known to prefer octahedral sites [46], and as Sn incorporation in bulk $Ga_2O_3$ is known to be sensitive to doping concentration and growth conditions, previous studies have reported segregation of Sn at the surface [32, 47, 48]. As mentioned previously, in HAADF-STEM we get a strong contrast for heavy Sn atoms compared to lighter Ga atoms [35]. Higher density of Sn dopants near the surface could therefore contribute to some of the large difference in contrast seen in bright octahedral Ga sites. More discussion on locating the Sn atoms can be found in the supplementary [Fig. S4-S5]. As such, surfactant effects of Sn and also Ge dopants [32, 47-49] may explain our observation of $\gamma$-$Ga_2O_3$ on the surface, however, the $\gamma$-phase surface layers of thickness less than 1 nm even in Si-doped $\beta$-$Ga_2O_3$ suggests that the temperature may play a crucial role in $\gamma$-phase formations, regardless of the types of dopants. However, more systematic study would be needed to confirm our suggestion.

Bhuiyan et. al. previously reported that the existence of the 90° rotated $\beta$-[010] structure from $\beta$-[001] caused by a high Al concentration in the $\beta$-$(Al_xGa_{1-x})_2O_3$ film could serve as an intermediate state inducing the $\gamma$-[110] structure [25]. However, all three of our doped samples show that the transition occurs rather directly, regardless of the $\beta$-[010] structure. One possible explanation as to why we see a direct phase transition, particularly on the surface of doped samples, has been suggested by the similarity of the oxygen sublattices of $\gamma$-$Ga_2O_3$ (110) when grown on $\beta$-(010) surface [50]. However, as we will see in the next section, $\gamma$-$(Al_xGa_{1-x})_2O_3$ does not form on the (010) surface of undoped $\beta$-$(Al_xGa_{1-x})_2O_3$. Therefore, we suggest that the dopants are also playing a crucial role in stabilizing the $\gamma$-phase. It has already been reported that doping with Mn or Fe can stabilize the growth of thin film $\gamma$-$Ga_2O_3$ [51, 52]. Here, Mn is known to occupy the tetrahedral sites in $\gamma$-$Ga_2O_3$ [51]. Similarly, Ge and Si may stabilize the $\gamma$-phase because they prefer to occupy tetrahedral sites as well. On the other hand, Sn doping is expected to have a different stabilization mechanism. While Sn is known to occupy octahedral sites, it is also reported that the Sn dopant can displace a Ga atom in the tetrahedral site, forming divacancy-interstitial complexes in $\beta$-$Ga_2O_3$ [53]. As Sn has a higher tendency for surface segregation, the presence of Sn dopants could generate a large population of mobile, interstitial Ga near the surface. As we discuss below, we suspect the $\gamma$-phase on the surface may be formed by the out-diffusion/precipitation of Ga interstitials.

As the samples are prepared by FIB which uses Ga ions to prepare a TEM lamella, questions may arise whether the $\gamma$-$Ga_2O_3$ on the surface of doped $\beta$-$Ga_2O_3$ is intrinsic to growth, or a result of ion beam damage. Since we have used the same sample preparation method for doped and alloyed samples, the presence of $\gamma$-phase only in doped film surfaces suggests that such a possibility is unlikely. Also, comparison of doped $\beta$-$Ga_2O_3$ films to that of undoped would be necessary to complete the understanding of dopants and the formation mechanism of the $\gamma$-phase, however the growth in current MBE systems in the presence of ion sources would result in unintentionally doped (UID) $\beta$-Ga$_2$O$_3$ films. The lowest UID Si-doped film achievable in our system had a doping density in the order of ~10$^{16}$ cm$^{-3}$, which still showed the formation of $\gamma$-phase surface layer (see Fig. S2).

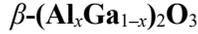

We now discuss $\beta$-(Al$_x$Ga$_{1-x}$)$_2$O$_3$ ($x$< 20%) epitaxial films grown on $\beta$-Ga$_2$O$_3$ substrates to compare the formation mechanism of $\gamma$-phase to that of the doped samples. Three $\beta$-(Al$_x$Ga$_{1-x}$)$_2$O$_3$ films were grown for an hour at 500 °C (A1), 600 °C (A2) and 700 °C (A3). Crystal growth schematics are shown in Fig. 6 (a). Sample A1 was grown on Sn-doped (010) $\beta$-Ga$_2$O$_3$ substrate while A2 and A3 were grown on Fe-doped (010) $\beta$-Ga$_2$O$_3$ substrates. The thicknesses of the $\beta$-Ga$_2$O$_3$ buffer layers are approximately half the thicknesses of the $\beta$-(Al$_x$Ga$_{1-x}$)$_2$O$_3$ films, as the growth rate of both $\beta$-(Al$_x$Ga$_{1-x}$)$_2$O$_3$ and UID Ga$_2$O$_3$ is similar. The films showed several orders of Pendellösung fringes indicating high quality, uniform films with smooth interfaces as shown in Fig. 6 (b). The thicknesses of the films were determined by fitting the measured XRD $2\theta$-$\omega$ scans using the Xrayutilities Python package [54], to be approximately 32 nm, 55 nm and 37 nm, respectively. The corresponding Al concentrations (A1-A3) of 12.0%, 15.5% and 19.4% were extracted using the separation of the (020) $\beta$-(Al$_x$Ga$_{1-x}$)$_2$O$_3$ and the (020) Ga$_2$O$_3$ Bragg angle peaks from a symmetric $2\theta$-$\omega$ scan, as described by Oshima et al. [55]. The decrease in growth rate (as shown in Table I.) and increase in Al content with the increase of substrate temperature is attributed to increase in desorption of the volatile Ga$_2$O suboxide with increasing substrate temperature, which leads to lower Ga concentration on the growth surface and an increased Al : Ga ratio [29]. AFM measurements in Fig. 6 (c) showed similar surface morphology for all three films with sub-nanometer roughness over a 2 $\mu$m×2 $\mu$m scan area. The RMS values for the films were 0.340 nm, 0.909 nm and 0.787 nm for the films A1-A3, respectively. All films showed elongated island growth in the [001] direction which is typically seen in Ga$_2$O$_3$ and (Al$_x$Ga$_{1-x}$)$_2$O$_3$ grown on (010) oriented surfaces by MBE [31, 47, 48].

The $\beta$-(Al$_x$Ga$_{1-x}$)$_2$O$_3$ samples were imaged by HAADF-STEM for atomic scale observations. The bulk of the film is in the $\beta$ phase aligned with the substrate. Consistent with an island growth/nucleation, sample A1 grown at 500 °C shows V-shaped regions between the ordered phase, and these contain polycrystalline inclusions as seen in Fig. 7 (a), (b). Although the TEM sample was cut along the (001) surface so the majority of the film is viewed along $\beta$-[001], inside the v-shaped regions we also see the coexistence of $\beta$-[010] and $\gamma$-[110] as in Fig. 7 (c). Extremely high density of Ga interstitials were observed (Fig.S6) in the $\beta$-phase near the transition to $\gamma$-[110].

TEM images of sample A2 grown at 600 °C display a reduced fraction of V-shaped defect regions, and these still include some $\gamma$-phase, at a frequency of approximately one region per 300 nm with a standard deviation of the mean of 48 nm in the lateral direction as shown in Fig. 7 (d). The regions showing the $\gamma$-Ga$_2$O$_3$ crystal structure are highlighted by red arrows in Fig. 7 (f). We also note that while there were few Ga interstitials overall, they are still notably present near the $\gamma$-phase interface.

Sample A3, grown at 700 °C is the most uniformly grown of the three samples, without any secondary phase inclusions. However, we observed high density of uniformly distributed Ga interstitials throughout the film as pointed out by yellow arrows in Fig. 7 (i).

As we were able to get an inclusion-free, uniformly grown film at 700 °C, we expect that higher growth temperatures should eliminate the formation of the $\gamma$-phase, as $\gamma$-Ga$_2$O$_3$ can be annealed out at higher temperatures [43]. We also focus on the fact that samples A1 and A2 have low Al concentrations (x<20%), yet they both showed inclusions of the $\gamma$-phase, while reference [25] did not observe inclusions of the $\gamma$-phase below (x<27%). This suggests that the $\gamma$-(Al$_x$Ga$_{1-x}$)$_2$O$_3$ formation could be more dependent on the growth conditions, rather than the Al composition.

The formation of $\gamma$-(Al$_x$Ga$_{1-x}$)$_2$O$_3$ in the growth of $\beta$-(Al$_x$Ga$_{1-x}$)$_2$O$_3$ is not yet fully understood. Bhuiyan et.al. [25] suggested that the Al substitution may create a strain to rotate $\beta$-Ga$_2$O$_3$ [001] into [010], and that the strain can then further trigger the transformation of $\beta$-Ga$_2$O$_3$ [010] into $\gamma$-Ga$_2$O$_3$ [110]. We also observe regions of $\beta$-(Al$_x$Ga$_{1-x}$)$_2$O$_3$ [010] phases nearby the $\gamma$-(Al$_x$Ga$_{1-x}$)$_2$O$_3$ [110], but given the authors' challenges in distinguishing the overlap of multiple faulted $\beta$-(Al$_x$Ga$_{1-x}$)$_2$O$_3$ [010] structures from that of $\gamma$-(Al$_x$Ga$_{1-x}$)$_2$O$_3$ [110], careful analysis still needs to be done to understand the extent of strain effects caused by Al substitution.

While sample A1 shows only a small fraction of $\gamma$-Ga$_2$O$_3$ inclusions uniformly distributed within the film, sample A2 shows localized growth of $\gamma$-Ga$_2$O$_3$ inside the V-shaped regions. The edges of the V-shapes form either 54° or 33-37° with the (010) surface, which are quite close to that of $(41\bar{2})$ and $(21\bar{1})$ planes in $\beta$-(Al$_x$Ga$_{1-x}$)$_2$O$_3$. These planes resemble the planes forming etch pits in bulk $\beta$-Ga$_2$O$_3$ previously reported by K. Hanada et.al. [56], suggesting that these planes have preferably lower surface energies. Along these planes, $\beta$-(Al$_x$Ga$_{1-x}$)$_2$O$_3$ [001] is either seen to rotate to $\beta$-[010] or transform into $\gamma$-(Al$_x$Ga$_{1-x}$)$_2$O$_3$ [110]. Especially along the $(21\bar{1})$ planes we see formation of $\gamma$-(Al$_x$Ga$_{1-x}$)$_2$O$_3$ as seen in Fig. 7 (e) and also in Fig. 8 showing an enlarged area inside the white boxed region in Fig. 7 (e). The (111) plane of the $\gamma$-phase is observed to start on the $(21\bar{1})$ plane of $\beta$-$(Al_xGa_{1-x})_2O_3$ (red arrows), which then rotates $107°$ to form a cluster of [110]-projected structures.

Although the atomic column intensities of a region in sample A2 shown in Fig. 8 are different from what we expect for a single $\gamma$-$Ga_2O_3$ [110]-projected structure, the ABF-STEM image clearly shows that the region is dominated by the $\gamma$-phase. The different atomic column intensity is due to several out-of-phase formations of $\gamma$-phase on the surface of $\beta$-$Ga_2O_3$ $(21\bar{1})$, leading to superposition of $\gamma$-$Ga_2O_3$ [110] sheets with different combinations of lattice shifts. The boxed area in red shows the lattice shifts happening in the ABF-STEM image. As a result, HAADF-STEM images taken in projection suffers from different atomic column intensity from that of the original single-grain lattice structure. Here again, we note that determining the crystal structure by depending on the atomic column intensity in HAADF-STEM images is not recommended as the image intensity can be affected largely by the superimposed crystal structure as well as the concentration of Al and the actual position of Al along the column that can affect the channeling of the electron probe. Rather, by ABF-STEM imaging we can identify the phases by obtaining the full crystal structure, with all lattice sites including the oxygen, visible.

## Correlation of Ga Interstitials to the Formation of $\gamma$-$Ga_2O_3$

Both in doped and alloyed films, we have frequently observed Ga interstitials in the $\beta$-phase structure near the region where transitions from $\beta$ to $\gamma$ would occur. Thus, for doped films, the interstitials were observed near the surface, while for alloyed samples A1 and A2, the interstitials were found inside the film near the phase-transition boundary. Sample A1 had excessive Ga interstitials as seen in Fig. S6. Although sample A2 had a few Ga interstitials, they were only found near the V-shaped regions. Sample A3 also had a high density of Ga interstitials, but as there were no $\gamma$-phase, it was rather uniformly distributed in the film.

Fig. 9 (a) presents the interstitial position in the crystal schematic of $\beta$-$Ga_2O_3$ in the crystal zones [010] and [001]. Here, blue Ga sites are tetrahedral Ga sites while yellow Ga sites are octahedral Ga sites. Both cross-section [001] and plan-view [010] TEM samples were prepared from Sn-doped $\beta$-$(Al_{0.15}Ga_{0.85})_2O_3$ (DSn). As seen in Fig. 9 (b), high density of Ga interstitials along a column is observed in plan-view [010]. When seen in cross-section [001], Fig. 9 (c) shows highly-occupied interstitial sites with intensity of interstitials comparable to that of the Ga atoms nearby. However, the reason for having such high density of Ga interstitials along a single column, and also the high density of occupied interstitial sites, i.e. why the interstitials cluster, is not yet clearly understood.

It is worth noting that $\beta$-Ga$_2$O$_3$ can support point-defect complexes where one cation interstitial atom is paired with two cation vacancies [53]. A defect complex having a vacancy with a neighboring Ga displaced from its tetrahedral site (to create a 2$^{nd}$ Ga vacancy) and moved to an interstitial site has the lowest formation energy, which can be denoted as a $2V_{Ga}^T$-Ga$_i$ defect complex. In the case of doped (Sn-, Ge-, Si-) $\beta$-Ga$_2$O$_3$ films, as dopants normally favor a different coordination than Ga, the defect complex may become $2V_{Ga}^T$-D$_i$ (D=dopants) complex instead. Thus, dopants provide one possible mechanism for increasing the population of interstitials.

To investigate the behavior of Ga interstitials and its relationship to $\gamma$-phase formation, we have performed ex-situ heating of TEM samples prepared from a Sn-doped $\beta$-(Al$_{0.15}$Ga$_{0.85}$)$_2$O$_3$. The samples used for ex-situ heating study were TEM lamellae having dimensions of 10 $\mu$m (length) × 5 $\mu$m (height) and 100 nm in average for the width. Such a small volume would pose restrictions on the migration of interstitials compared to the bulk. This enables us to detect the changes or movements of the interstitials more easily. Also, the interstitials in our experiment would experience more migration towards the surface than in bulk, leading to much larger formation of the $\gamma$-phase if interstitials do play a role.

After acquiring HAADF-STEM images at room temperature, we heated the samples in an Ar atmosphere up to 200 °C and imaged it again. After seeing no changes, we once again heated up to 400 °C in Ar, then imaged it again after cooling. For accurate comparison, the same region was observed in each case. Fig. 10 shows the resulting HAADF-STEM images. At room temperature we observed the surface was terminated with the uppermost unit cell in the $\gamma$-phase as shown in Fig. 10 (a). After being heated up to 200 °C as in Fig. 10 (b), the surface in the same region is essentially unchanged. However, after the 400 °C anneal as shown in Fig. 10 (c), a 5-8 nm-thick layer of $\gamma$-phase had formed which was grown above the original surface. More images are shown in Fig. S7.

The likely origin of this extra $\gamma$-phase layer are the interstitials observed in the bulk of the $\beta$-phase that have migrated towards the surface. The more open $\gamma$-phase is likely to allow faster diffusion and growth kinetics at intermediate temperatures. Fig. 4 suggests that by 650 °C, the $\beta$-phase dominates again. To obtain an estimate of changes in the interstitial density we turn to some subtle details of our STEM imaging. In STEM, we use a finely focused probe to scan across the sample in a raster pattern to obtain images. However, imaging interstitials under an electron microscope with a beam energy of 300 keV can be challenging as the energy of the electron beam is sufficient to displace the interstitials, causing them to occasionally hop around from one interstitial site to another. We maintained the same voltage, current, and dwell time each time we imaged the samples, so that the energy delivered to an interstitial site remained the same. We were not able to see much evidence of hopping of the interstitials in the sample before and after being heated up to 200 °C. However, frequent hopping was observed in the sample after being heated up to 400 °C.

In Fig. 9 (b) we have seen that at room temperature there is a high density of interstitials along a column as well as a high density of occupied interstitial sites in the film, shown in Fig. 9 (c). As TEM sees atoms in projection, if the interstitials are hopping to a nearby interstitial site, where other interstitials sites in the same column are already occupied, the contrast change would be low and we would not be able to observe the hopping clearly. Even if one or two interstitials left the original position, we would still detect the remaining interstitials at the previous site, so the change in contrast would be small. Imaging the sample after heating up to 400 °C still showed similar fraction of occupied columns, however the intensity fluctuations from hopping was much easily observed. This implies that after the 400 °C anneal, we have a reduced density of interstitials along any given column.

As the growth of the $\gamma$-phase occurred when the number of interstitials decreased (i.e. 400 °C), and that it grew above the original surface, rather than as a conversion of existing $\beta$-phase, we suspect that the $\gamma$-phase was constructed from the out-diffusion and coalescence of interstitials. The growth of $\gamma$-phase at ~400 °C suggests some care must be taken to control local heating to avoid the possibility of device degradation while in operation. Although a typical operating temperature of a device is around 100-200 °C, the low thermal conductivity of $\beta$-$Ga_2O_3$ in combination with high voltage, high current conditions for extended periods could raise the temperature to a point where $\gamma$-phase growth from interstitial diffusion and coalescence could occur at interfaces like metal contacts. Therefore, all effort should be made to avoid the formation of the $\gamma$-phase.

## CONCLUSIONS

In summary, we have observed $\gamma$-phase inclusions in MBE-grown Ge-, Si- doped $\beta$-$Ga_2O_3$ films, Sn-doped $\beta$-$(Al_{0.15}Ga_{0.85})_2O_3$, and series of $\beta$-$(Al_xGa_{1-x})_2O_3$ films grown at different temperatures. The inclusion of $\gamma$-phase in most samples shows that it is a widely occurring defect during the growth of $\beta$-$Ga_2O_3$. The formation of a surface $\gamma$-phase is found in doped samples, especially at growth temperatures below 650 °C. We also observed $\gamma$-$(Al_xGa_{1-x})_2O_3$ formation between nucleating islands of $\beta$-$(Al_xGa_{1-x})_2O_3$ grown at temperatures below 650 °C. From the trend seen in all our doped $\beta$-$Ga_2O_3$, and $\beta$-$(Al_xGa_{1-x})_2O_3$ films, increasing the substrate temperature leads to a reduction in $\gamma$-phase inclusions and improved film crystalline quality and uniformity. The appearance of the $\gamma$-phase at lower temperatures suggests that $\gamma$-phase would be the early forming phase in terms of kinetics. This idea is further strengthened by our observation of the $\gamma$-phase growth above the original surface of Sn-doped $\beta$-$(Al_xGa_{1-x})_2O_3$ when heated up to a relatively low temperature of 400 °C, likely formed from the out-diffusion/precipitation of interstitials, again suggesting that the more-open $\gamma$-phase was kinetically favored over the more thermodynamically-stable $\beta$-phase.

## SUPPLEMENTARY MATERIAL

See supplementary material for more information on HAADF-STEM images of doped and alloyed samples.

## DATA AVAILABILITY

The data that supports the findings of this study are available within the article and its supplementary material. Raw data files are available on request from the corresponding author.

## ACKNOWLEDGEMENTS

This work was primarily supported by the Cornell/AFOSR ACCESS center of excellence (No. FA955018-1-0529). This work made use of the Cornell Center for Materials Research (CCMR) Shared Facilities, which are supported through the NSF MRSEC Program (No. DMR-1719875). The FEI Titan Themis 300 was acquired through Grant No. NSF-MRI-1429155, with additional support from Cornell University, the Weill Institute, and the Kavli Institute at Cornell. S. M. and T. J. A. thank for the support of OSD seeding grant entitled "150 Volt Ultra-wide Bandgap High-Efficiency RF Amplifier Technology." We also thank Y. Yao in Cornell University for the help with ex-situ heating using their heating furnace.

| Sample Label | Sample | $T_{sub}$ (°C) | Ga BEP (Torr) | Al BEP (Torr) | $O_2$ (Flow, sccm/RF Power, W/ $P_{GM}$, Torr) | Growth Rate (nm/hr) | Al% from XRD | AFM 2x2 µm² RMS (nm) |
|---|---|---|---|---|---|---|---|---|
| DG1 | Ge-$Ga_2O_3$ | 550 | $6.00 \times 10^{-8}$ | N/A | $1.55 / 300 / 1.50 \times 10^{-5}$ | 48.5 | N/A | 0.71 |
| DG2 | Ge-$Ga_2O_3$ | 600 | $2.05 \times 10^{-8}$ | N/A | $0.57 / 250 / 1.00 \times 10^{-5}$ | 38 | N/A | 0.80 |
| DG3 | Ge-$Ga_2O_3$ | 650 | $6.00 \times 10^{-8}$ | N/A | $1.55 / 300 / 1.50 \times 10^{-5}$ | 50.5 | N/A | 1.51 |
| DSn | Sn-$(Al_{0.15}Ga_{0.85})_2O_3$ | 550 | 770 °C (Cell T) | 876 °C (Cell T) | 3 sccm $O_3$ flow rate / $SnO_2$ 750 °C Cell T | 167 | 15 (From RBS) | 0.50 |
| DSi | Si-$Ga_2O_3$ | 650 | $3.10 \times 10^{-8}$ | N/A | $1.40 / 250 / 2.28 \times 10^{-5}$ | 86.5 | N/A | 0.47 |
| A1 | $(Al_{0.12}Ga_{0.88})_2O_3$ | 500 | $5.00 \times 10^{-9}$ | $1.00 \times 10^{-9}$ | $0.57 / 250 / 1.00 \times 10^{-5}$ | 32 | 12.0 | 0.34 |
| A2 | $(Al_{0.16}Ga_{0.84})_2O_3$ | 600 | $2.06 \times 10^{-8}$ | $3.84 \times 10^{-9}$ | $0.57 / 250 / 9.52 \times 10^{-6}$ | 55 | 15.5 | 0.91 |
| A3 | $(Al_{0.19}Ga_{0.81})_2O_3$ | 700 | $2.06 \times 10^{-8}$ | $3.84 \times 10^{-9}$ | $0.57 / 250 / 9.53 \times 10^{-6}$ | 38 | 19.2 | 0.79 |

**Table I:** Summary of the growth conditions for doped and undoped $\beta$-$(Al_xGa_{1-x})_2O_3$ samples discussed in this study. DG1 and DG3 were grown using a Veeco GEN Xcel PAMBE system in AFRL, while the other samples were grown in a Veeco Gen930 MBE system at Cornell university. The growth rate was extracted from SIMS for DG1 and DG3 and from XRD for all other samples.

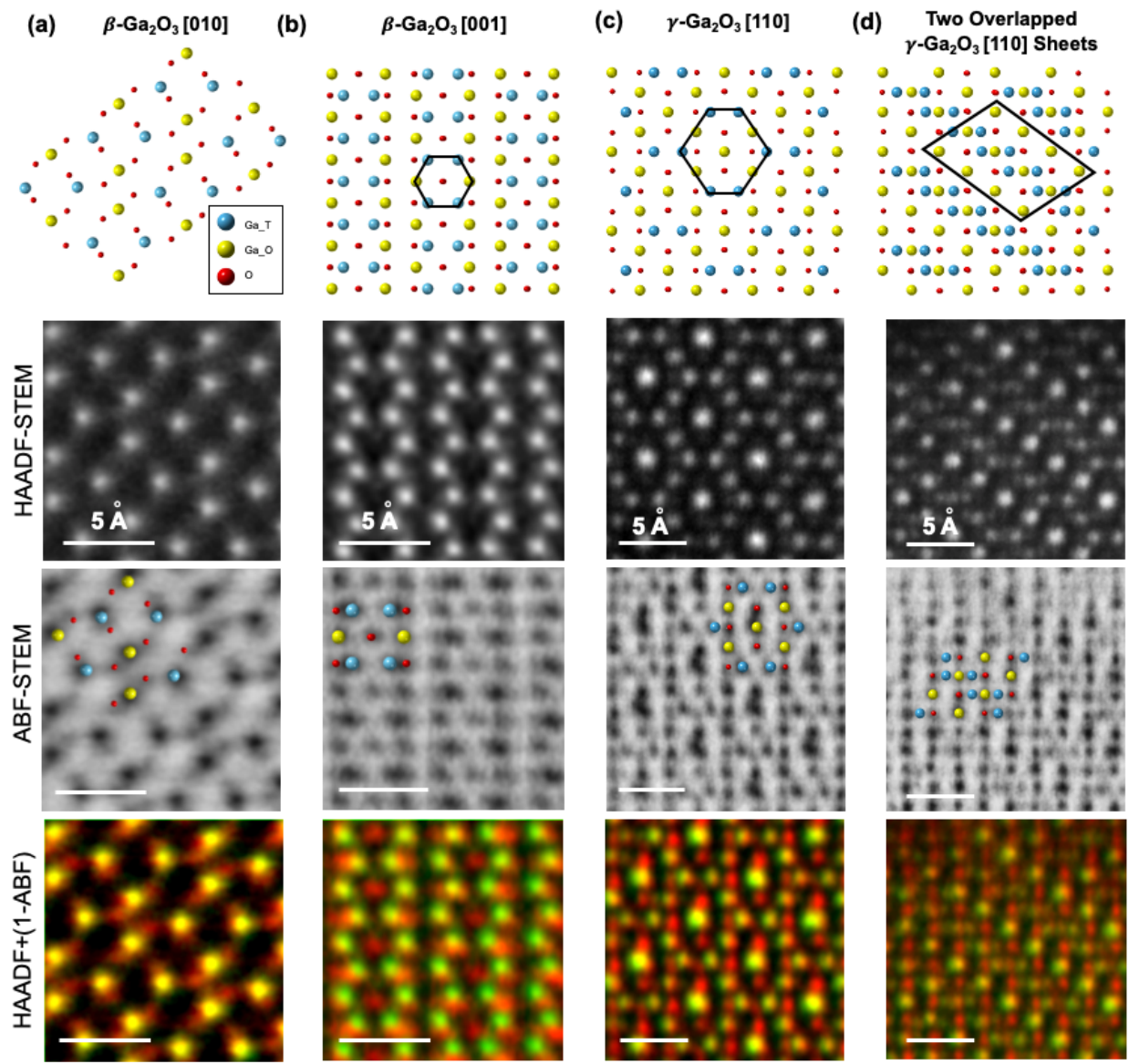

**Figure 1**: Crystal structures of $\beta$ and $\gamma$-phases in different crystal zones. (a) Crystal structure of $\beta$-Ga$_2$O$_3$ in [010] crystal zone. The blue atoms and yellow atoms each correspond to the tetrahedral Ga $T$ and octahedral Ga $O$ sites respectively. (b) Crystal structure of $\beta$-Ga$_2$O$_3$ in [001] crystal zone. It forms a hexagonal crystal structure consisting of four tetrahedral sites and two octahedral sites. (c) $\gamma$-Ga$_2$O$_3$ structure in [110] crystal zone. The projection of Ga atoms again resembles a hexagonal pattern as in (b), instead it now has four tetrahedral sites and six octahedral sites in the bigger hexagon, with one of the octahedral sites sitting in the middle of the hexagon. (d)

Superimposed crystal structure of two misaligned [110] $\gamma$-Ga$_2$O$_3$ creating a ladder-like lattice structure. Below the model structures are the HAADF and ABF STEM images for each of these zone axes. The HAADF contrast is dominated by, and brightest on, the Ga sites. The ABF contrast is opposite, darkest on the Ga sites, but also shows the oxygen sites. On the bottom row are false color overlays of the HAADF in the green channel, and the ABF image with contrast inverted in the red channel. The false color map thus reveals the full crystal structure, with oxygen atoms as red and Ga as green.

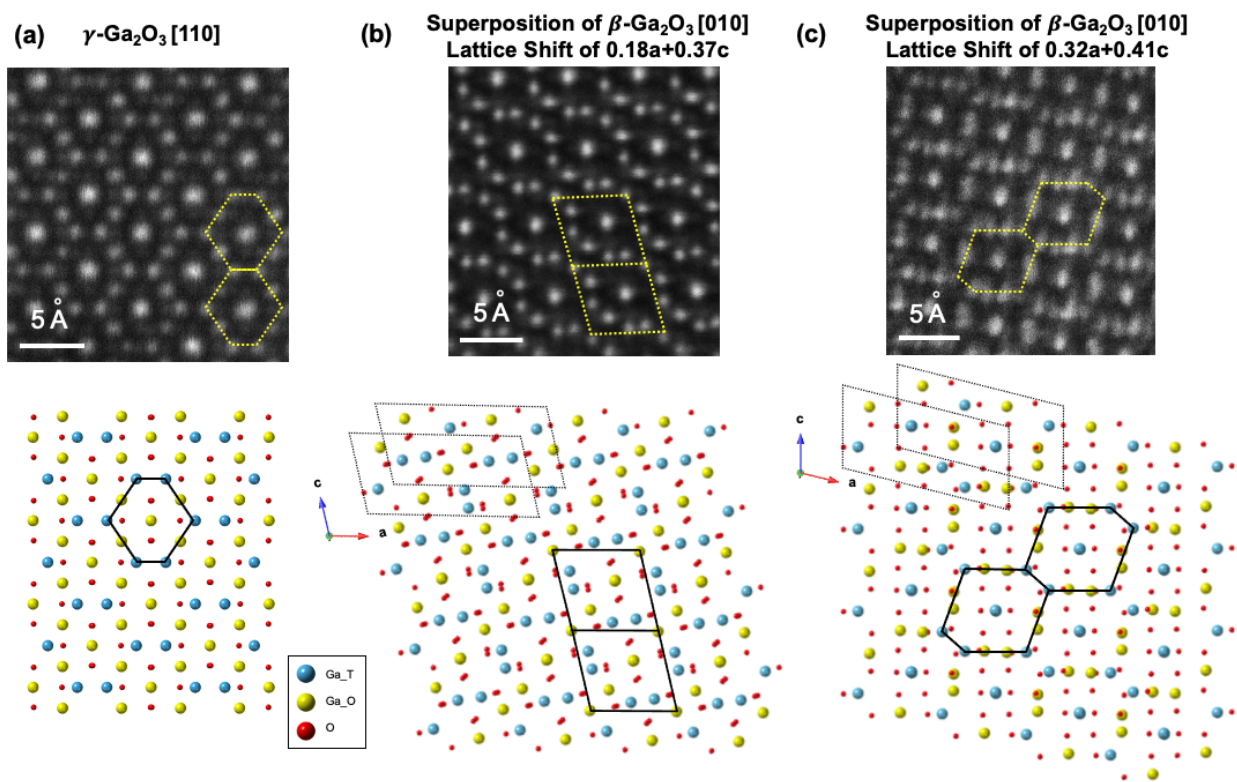

**Figure 2**: Comparison of the $\gamma$-Ga$_2$O$_3$ crystal structure to two differently-superimposed $\beta$-Ga$_2$O$_3$ lattice structures separated by a stacking fault by aberration-corrected STEM images and crystal models. The 2D-projected repeat unit for each crystal structure is outlined by yellow dotted lines on the HAADF-STEM images and black solid lines in the crystal model. (a) $\gamma$-Ga$_2$O$_3$ is characterized by edge-sharing hexagonal honeycomb structure where the four side-edges have three Ga atoms arranged in a straight line. (b) shows the superimposed lattice structure after a net lattice shift of 0.18a and 0.37c along the a and c direction respectively. Black dotted lines outline the unit cell of $\beta$-Ga$_2$O$_3$. Although the outline of the crystal structure is superficially similar to hexagonal $\gamma$-Ga$_2$O$_3$, this crystal structure has rounded side-edges and the 2D-projected repeat unit is not hexagonal. (c) shows the superimposed lattice structure after a combined lattice shift of 0.32a and 0.41c (or 0.18a and 0.58c) along the a and c direction respectively. This crystal structure has four atoms in the side-edges and has a distorted hexagonal repeat unit in projection.

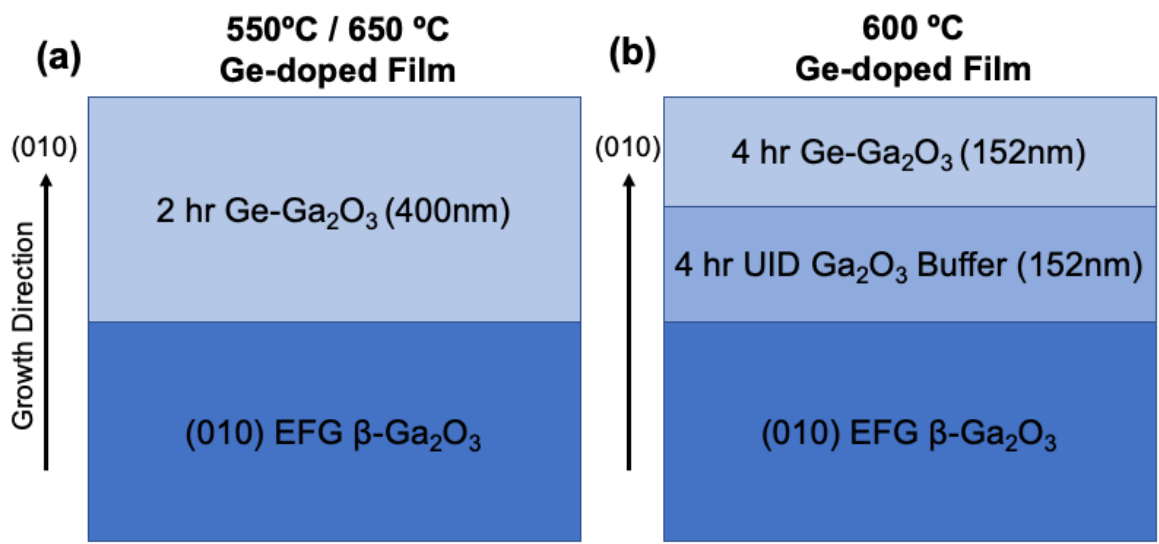

**Figure 3**: Crystal growth schematics for Ge-doped $\beta$-Ga$_2$O$_3$ samples grown at different temperatures. (a) Ge-doped $\beta$-Ga$_2$O$_3$ films were grown at 550 °C and 650 °C on Fe-doped $\beta$-Ga$_2$O$_3$ substrates. The film thickness is 400 nm. (b) Film grown at 600°C have a buffer layer with a thickness same as that of the film to prevent the diffusion of Fe dopants from the substrate.

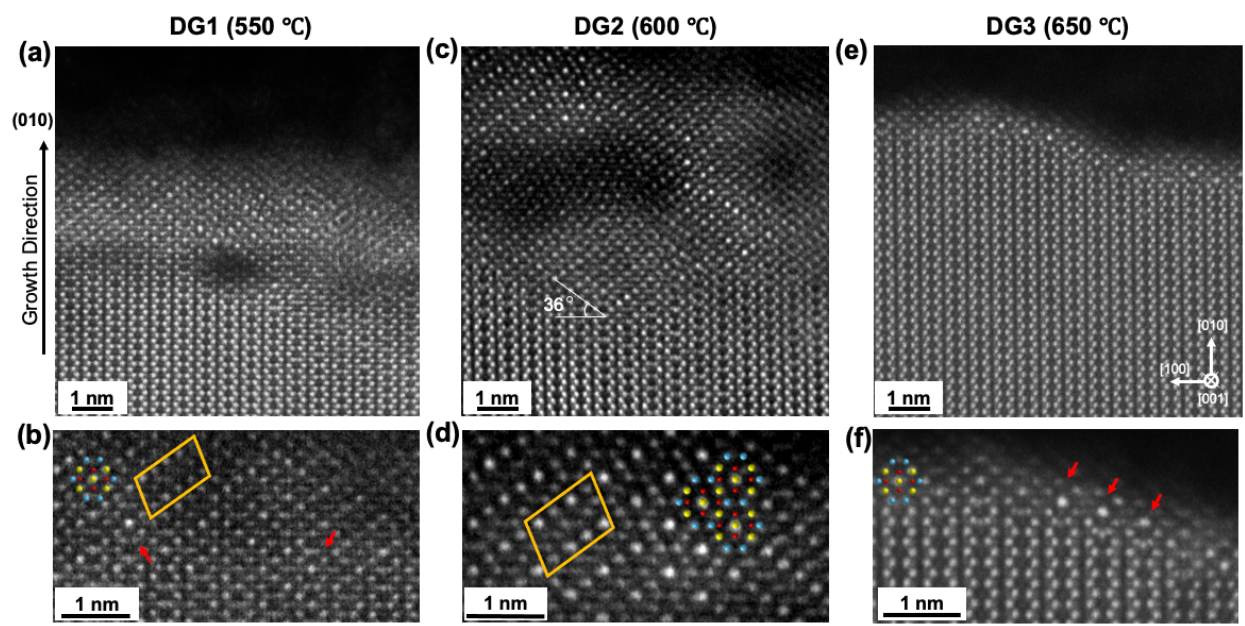

**Figure 4**: Comparison of Ge-doped $\beta$-Ga$_2$O$_3$ samples grown at different temperatures. All three samples show $\gamma$-phase formations at the film surface. (a)-(b), (c)-(d), (e)-(f) shows films grown at 550 °C (DG1), 600 °C (DG2), and 650 °C (DG3) respectively. Enlarged surfaces are shown in (b), (d), (f). A distinguishable $\gamma$-Ga$_2$O$_3$ lattice structure can be observed on top of $\beta$-Ga$_2$O$_3$ in DG1 and DG2, as in (a) and (c), which is further enlarged to (b) and (d) to show the crystal structures including the misaligned $\gamma$ lattice structure discussed in Fig. 1 (d). Red arrow shows the $\gamma$-Ga$_2$O$_3$ formation occurring inside a column of $\beta$-Ga$_2$O$_3$. Film grown at 650 °C (DG3) in (e), (f) shows an onset of $\gamma$-phase (~1nm) by the nucleation at bright octahedral Ga sites. All three samples show Ga interstitials near the $\gamma$-phase, which can be found in (a), (c) and (e).

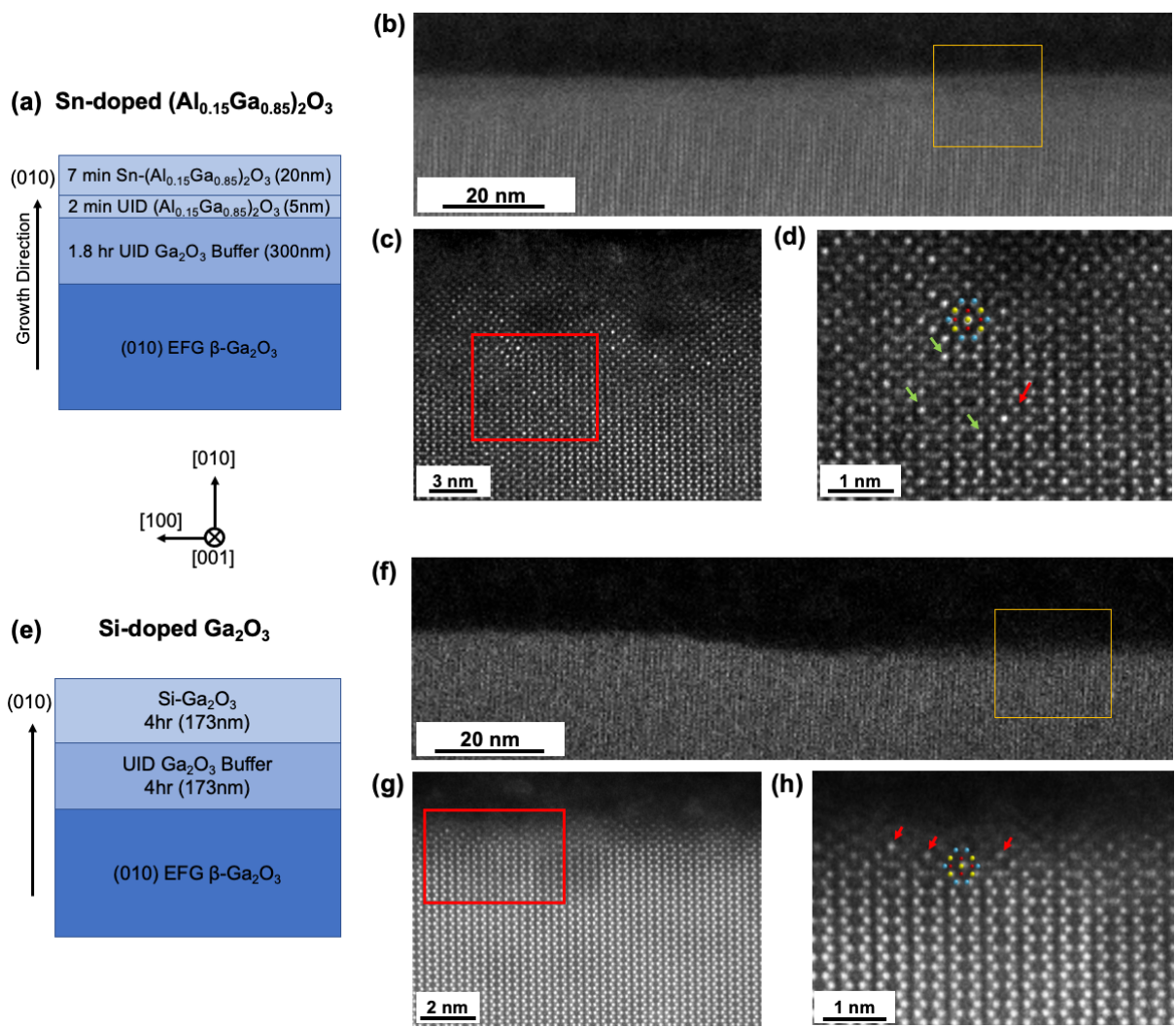

**Figure 5**: Sn-doped $\beta$-$(Al_{0.15}Ga_{0.85})_2O_3$ (DSn) sample and Si-doped $\beta$-$Ga_2O_3$ (DSi) sample showing $\gamma$-phase formations at the film surface. (a) Growth schematic of DSn grown at 550 °C. This sample also shows (b) distinct surface layer of $\gamma$-$(Al_xGa_{1-x})_2O_3$ which is enlarged in (c). (d) shows inclusion of gamma phase on the surface as well as inside the $\beta$-phase crystalline structure shown by a red arrow. Green arrows point to octahedral Ga atomic sites that have higher contrast compared to other Ga sites, suggesting they are Sn-dopants occupying the octahedral sites. DSi was grown at 650°C with a buffer layer as in (e). (f) Low magnification image does not show distinct $\gamma$-$Ga_2O_3$ layer on the surface. (g) and (h) shows an onset of $\gamma$-phase by the nucleation at bright octahedral Ga sites. This observation is similar to Ge-doped $\beta$-$Ga_2O_3$ grown at 650°C (DG3).

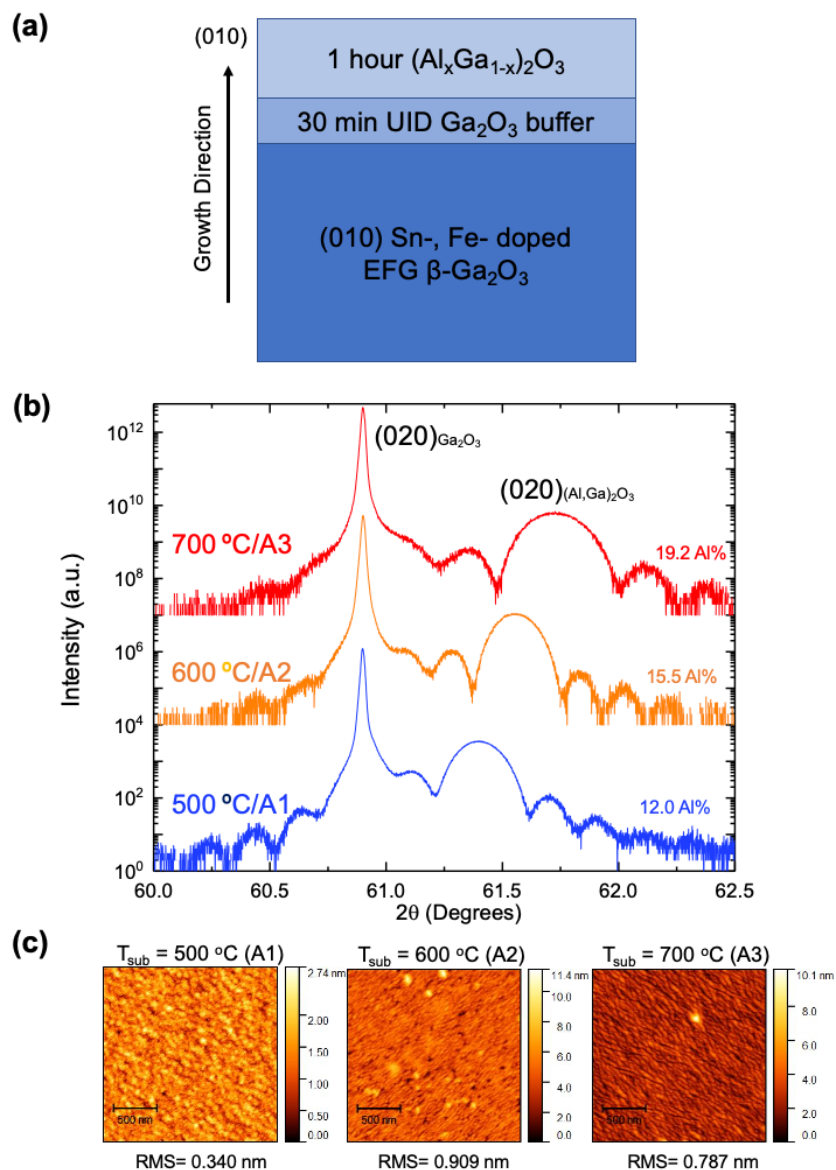

**Figure 6**: (a) Crystal growth schematics for series A1-A3. A1 was grown on Sn-doped (010) $\beta$-Ga$_2$O$_3$ substrate while A2 and A3 were grown on Fe-doped (010) $\beta$-Ga$_2$O$_3$ substrate. (b) 2$\theta$-$\omega$ XRD scans show the shift in the $\beta$-(Al$_x$Ga$_{1-x}$)$_2$O$_3$ (020) peak as T$_{sub}$ is changed from 500-700 °C as well as a change in thickness for the film grown at 700 °C. (c) Comparison of 2×2 $\mu$m$^2$ AFM scans. Films are smooth at all growth temperatures but typical surface morphologies showing elongated islands in the [001] direction becomes more prominent at T$_{sub}$=600 °C and higher.

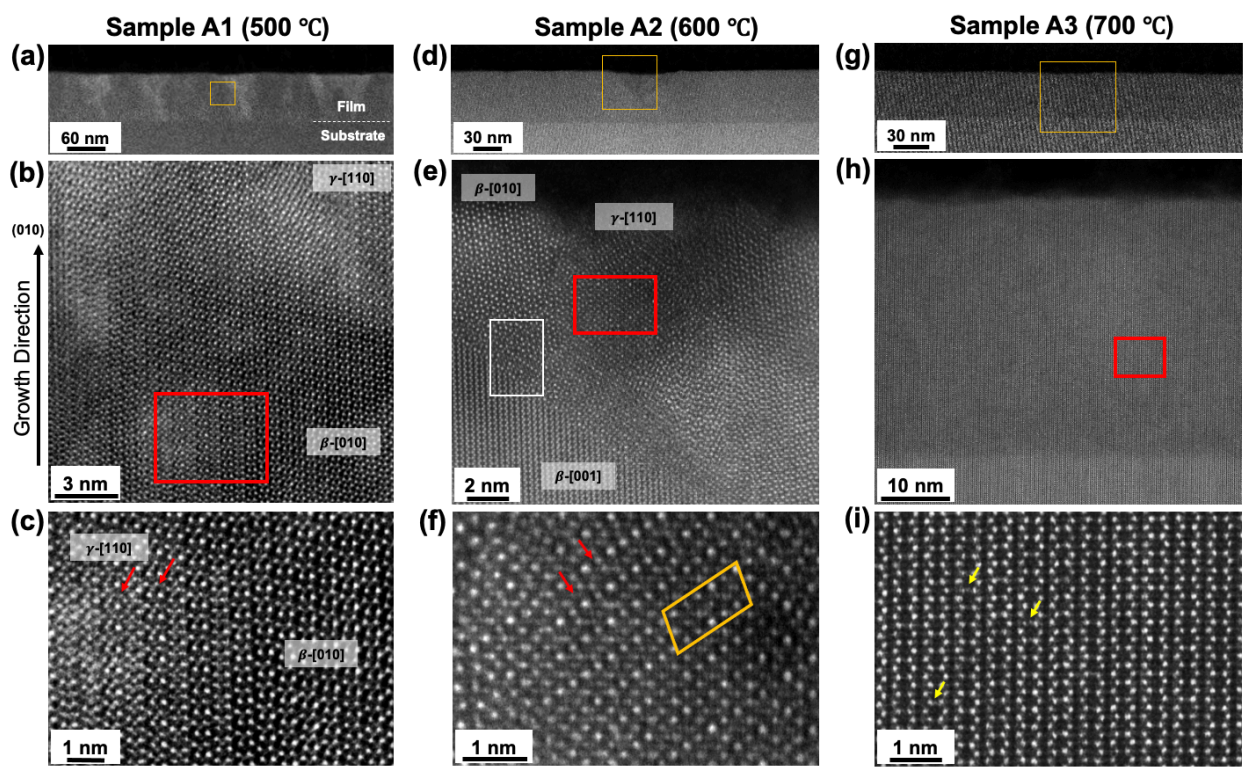

**Figure 7**: Samples A1-A3 (500-700 °C) grown with increasing temperature reveal different characteristics in TEM images. (a)-(c), (d)-(f), (g)-(i) shows A1, A2, and A3, respectively. Boxed region in (a), (d), (g) enlarges to (b), (e), (h) then to (c), (f) and (i). Overview of Sample A1 (500 °C) in (a) shows large regions of polycrystalline inclusions. Enlarged image in (b) and (c) shows mixture of $\beta$ [010] and $\gamma$ [110]-crystal zones, where the actual crystal zone for this projection should be $\beta$ [001]. Sample A2 (600 °C) in (d) shows a cleaner film than in (a), however, it has V-shaped defect regions. (e) shows inside the V-region where the formation of both $\beta$ [010] and $\gamma$ [110] phases have occurred. The region inside the white box will be discussed later in Fig. 8. (f) Enlarged area inside the V-region shows the crystal structure of $\gamma$-Ga$_2$O$_3$ [110] and superimposed $\gamma$-Ga$_2$O$_3$ [110]. Sample A3 (700 °C) in (g) and (h) shows a uniformly grown film. (i) Although the film does not suffer from $\gamma$-phases, it shows high density of Ga interstitials. Red arrows in (c) and (f) point to $\gamma$ [110] crystal structure while yellow arrows in (i) point to Ga interstitials.

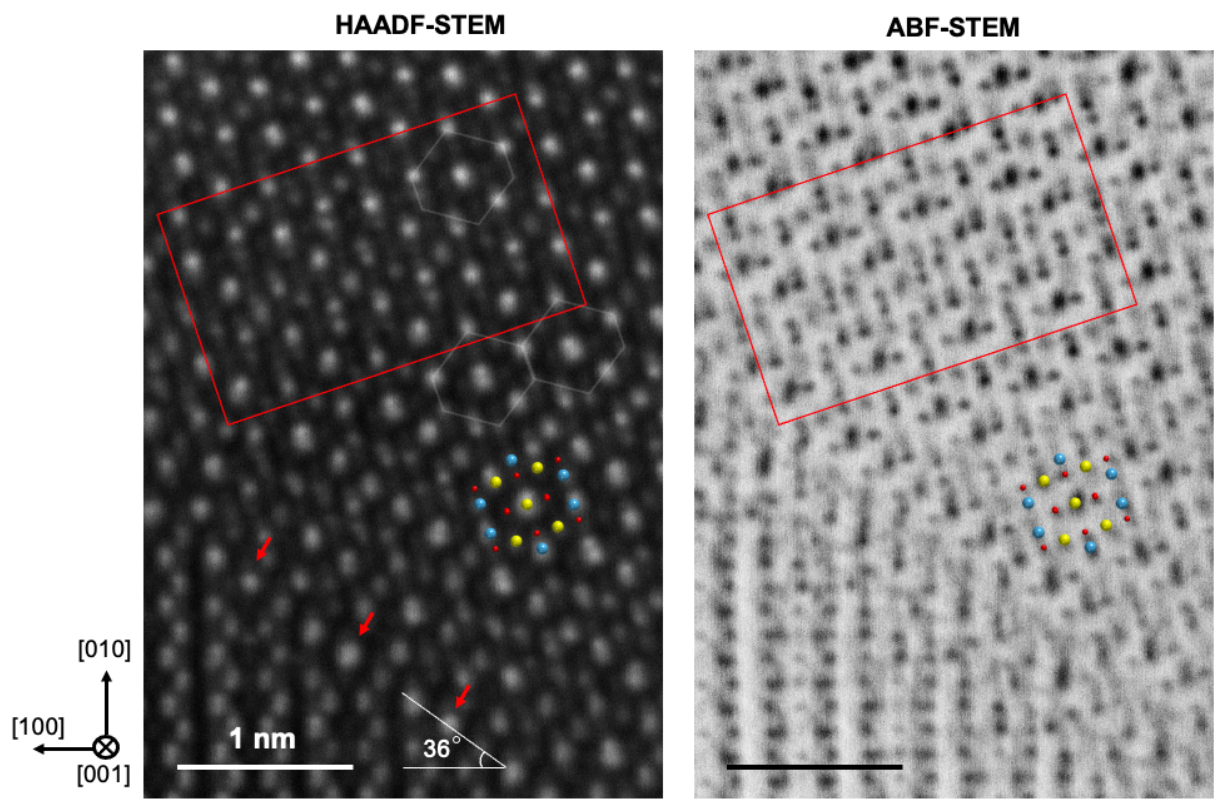

**Figure 8**: Observation of γ-(Al$_x$Ga$_{1-x}$)$_2$O$_3$ [110] formation directly on Sample A2 (600 °C), β-(Al$_{0.16}$Ga$_{0.84}$)$_2$O$_3$ [001]. This shows an area enlarged from the white box in Fig. 7 (e). The bottom right corner is slightly out of focus in ABF-STEM image due to sample tilt. The lower left side shows the crystal structure of β-Ga$_2$O$_3$ [001]. γ-(Al$_x$Ga$_{1-x}$)$_2$O$_3$ is observed to grow directly on the angled surface of sample A2, as shown by red arrows. Because the nucleation of γ-phase happens out-of-phase at different sites, several layers of γ-(Al$_x$Ga$_{1-x}$)$_2$O$_3$ [110] with different lattice shifts becomes superimposed as can be seen inside the red boxed area. This results in different intensities of Ga atomic sites in γ-(Al$_x$Ga$_{1-x}$)$_2$O$_3$. White outline of the γ-phase repeat unit is overlaid in the HAADF-STEM image to aid comparison of atomic intensities with that shown in Fig. 1 (c).

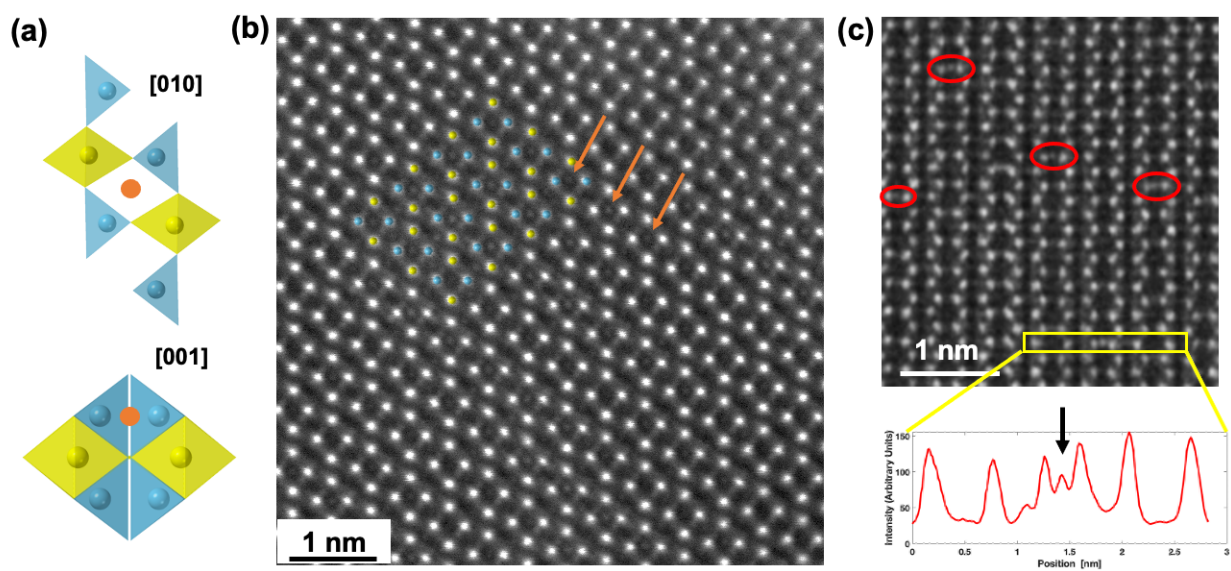

**Figure 9**: High densities of Ga interstitials observed in Sn-doped $\beta$-(Al$_{0.15}$Ga$_{0.85}$)$_2$O$_3$ (DSn) sample. (a) Crystal structure of $\beta$-Ga$_2$O$_3$ in [010] and [001] crystal zone showing interstitial positions as shown by the orange atom. Blue atoms correspond to tetrahedral Ga while yellow atoms are octahedral Ga. (b) High densities of interstitials are observed in Plan-view [010] DSn. Arrows are overlaid for guide to the eye. (c) In cross-section [001] of DSn we can see the Ga interstitials in red-circled areas. The line profile of an atomic row at the bottom of the figure shows considerable contrast from interstitial site compared to that of the two Ga columns nearby – i.e. there are multiple interstitials viewed in projection along the column.

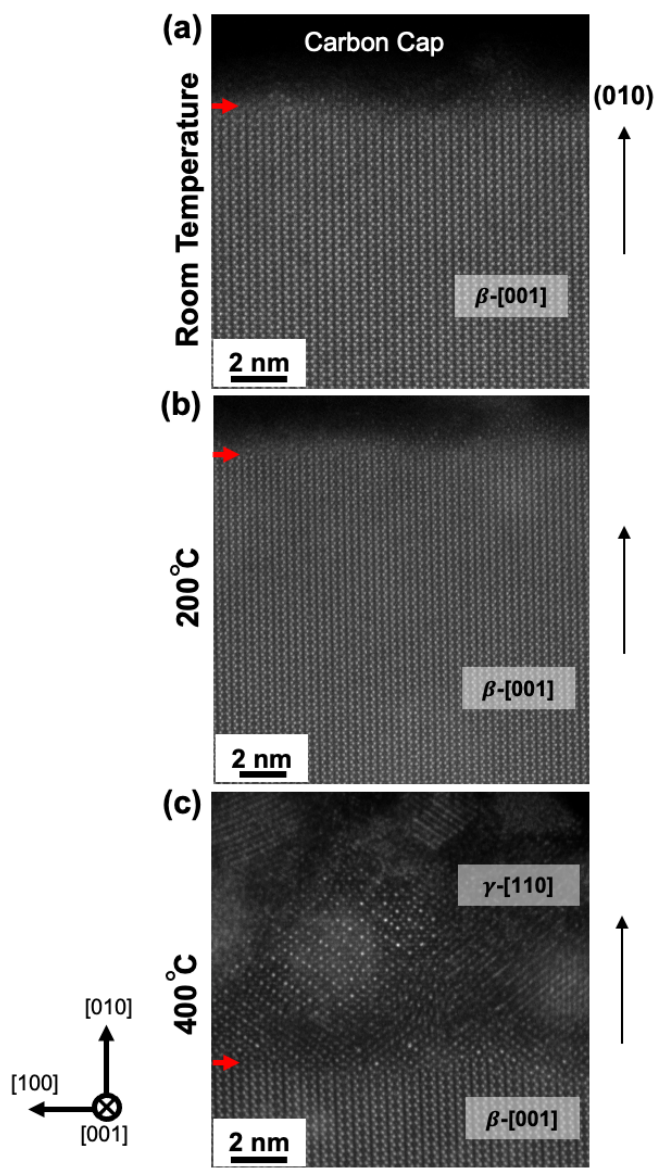

**Figure 10**: Ex-situ heating results of the same region of a Sn-doped $\beta$-$(Al_{0.15}Ga_{0.85})_2O_3$ (DSn) TEM cross-section. Both (a) the surface at room temperature and (b) the surface after being heated up to 200 °C shows only a $\gamma$-like surface termination. (c) After heating up to 400 °C, we see the formation of a thick (~5-8 nm) $\gamma$-phase region above the original surface. Red arrows indicate the approximate location of the original surface -i.e. the $\gamma$-phase has grown on top of the $\beta$-phase, presumably from the out-diffusion of interstitials, some of which are still present near the interface. Note the similar small island of $\gamma$-phase on the right in each image.

# SUPPLEMENTAL MATERIAL

Fig. S1 shows the surfaces of Sn-doped $\beta$-$(Al_{0.15}Ga_{0.85})_2O_3$ (DSn) and Ge-doped $\beta$-$Ga_2O_3$ (DG2). We have prepared several samples of DSn from different regions in the sample to see if the formation of $\gamma$-layers is prevalent in the sample. As shown in Fig. S1(a), at some regions we were not able to see any $\gamma$-phase, but most of the areas suffered from $\gamma$-inclusions. Fig. S1(b) and (c) shows that there is a surface layer on DG2. However, not all surface layers are $\gamma$-phase. We have also looked at UID $\beta$-$Ga_2O_3$ sample grown at 650 °C with ~$10^{16}$ cm$^{-3}$ Si dopant concentration as a control sample, which is the lowest achievable dopant concentration in our system. Fig. S2 shows images of UID $\beta$-$Ga_2O_3$ with a thin $\gamma$-phase at the surface, similar to Ge-doped $\beta$-$Ga_2O_3$ (DG3) sample.

Sn-dopants are known to segregate on the surfaces as it is sensitive to growth conditions. Fig. S3 shows DSn in the [001] crystal zone. In Fig. S3 (a), brighter octahedral Ga sites can be seen on the surface but also inside the film. We also observe high density Ga interstitials which is shown in Fig. S3 (b). In scanning transmission electron microscopy (STEM), selecting the detector geometry to collect electrons scattered at high angles, provides a signal dominated by Rutherford scattering from the nuclear potential of the targeted atoms. As a result, the atomic-number (Z) sensitivity of high angle annular dark field (HAADF) signal roughly scales as $Z^{1.7}$, making it easier to distinguish heavy Tin atoms from the lighter Gallium atoms. However the contrast obtained in an image is from the integrated, scattered signal from the electron beam that has propagated through several unit cells. As one Ga atom scatters as strongly as four Al atoms, the observable fluctuation in Al content by Rutherford scattering is less than the change in contrast due to the addition or removal of a single Ga site. To be specific, the contrast ratio of atomic columns having exactly 15% of Al to a pure Ga atomic column will have reduced contrast of approximately 10% as shown in Fig. S4. Actual contrast ratio will fall below 10% due to Poisson fluctuations of Al atoms in the column and other artifacts such as scanning noise during the acquisition of TEM images. Thus, we can safely and directly determine the inclusion of Sn atoms without the effect of Al fluctuations on the Ga sites, in the Sn-doped $\beta$-$(Al_{0.15}Ga_{0.85})_2O_3$ layer. On the other hand, as a single Sn atom would scatter as strongly as 2-3 Ga atoms, even low concentrations of Sn can give much higher contrast ratio. The first datapoints for different tin concentrations in Fig. S4 are the contrast dominated by one single Sn atom. The following second, third, and fourth datapoints correspond to contrast dominated by two, three and four Sn atoms respectively. As our actual Sn concentration is well below 1%, the contrast difference that will be shown in Fig. S5 is largely due to a single Sn atom. Such contrast can be enhanced once more by preparing a thin specimen by using Focused Ion Beam (FIB) which generally produces 10-20nm thick samples, to benefit from the tendency of increasing contrast with thinner samples.

Inside a single unit cell of $\beta$-$Ga_2O_3$, there are two crystallographically-inequivalent Ga atom sites, namely tetrahedral and octahedral Ga sites [57]. First principles calculations performed by Varley et al. [46] and S. Lany [58] indicate an energetic preference for Sn to preferentially substitute Ga at the octahedral site. Fig. S5 (a) shows the two sites for Ga: the octahedral sites are red, and the tetrahedral sites are in blue. The structural overlay of Fig. S5 (a) on the HAADF-STEM images taken from (b) the bulk and from (c) a thinner surface region of Sn-doped $(Al_{0.15}Ga_{0.85})_2O_3$ sample shows that the Sn atoms indeed occupy only the octahedral sites in (c).

Fig. S6 shows additional images of sample A1. In Fig. S6 (a), the film reveals coexistence of $\beta$ [001] and [010] domains. At the boundary of those two domains, Ga interstitials can be observed. (b) shows a region close to the domain boundaries that shows extremely high density of Ga interstitials. Fig. S7 shows two different cases of DSn after heating up to 400 ℃. Fig. S7 (a) and (b) shows the growth of the $\gamma$-phase after heating up a sample that had minimal $\gamma$-phase surface layer. Fig. S7 (c) and (d) shows a stabilized $\gamma$-phase of a sample that already had $\gamma$-phase prior to heating. The stability was compared by observing the beam damage on the surface $\gamma$-phase layer before and after ex-situ heating.

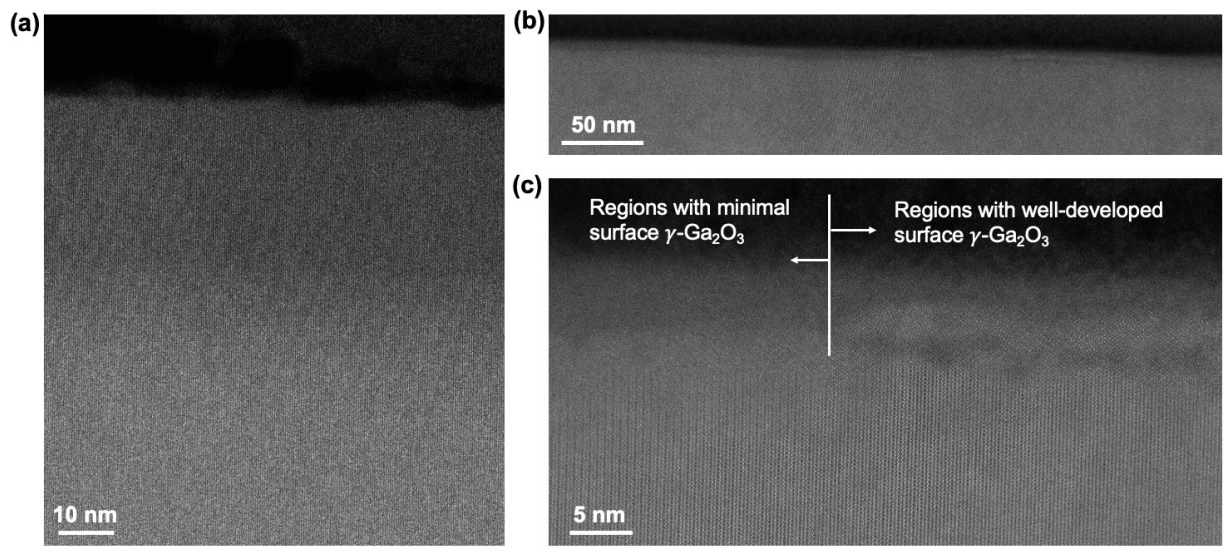

**Figure S1:** Region-dependent surface layers in Sn-doped $\beta$-(Al$_{0.15}$Ga$_{0.85}$)$_2$O$_3$ (DSn) and Ge-doped $\beta$-Ga$_2$O$_3$ (DG2). (a) TEM image of DSn taken at a region that does not show any surface layers or $\gamma$-phase inclusion. (b) Large field of view image of DG2 shows a distinct surface layer, where at some regions the $\gamma$-phase is present. The remainder of the surface layer is amorphous or highly disordered. (c) Enlarged image of DG2 also show that the phase separation does not happen at every region on the surface. The left side of the surface layer is amorphous, while the right contains a layer of the $\gamma$-phase.

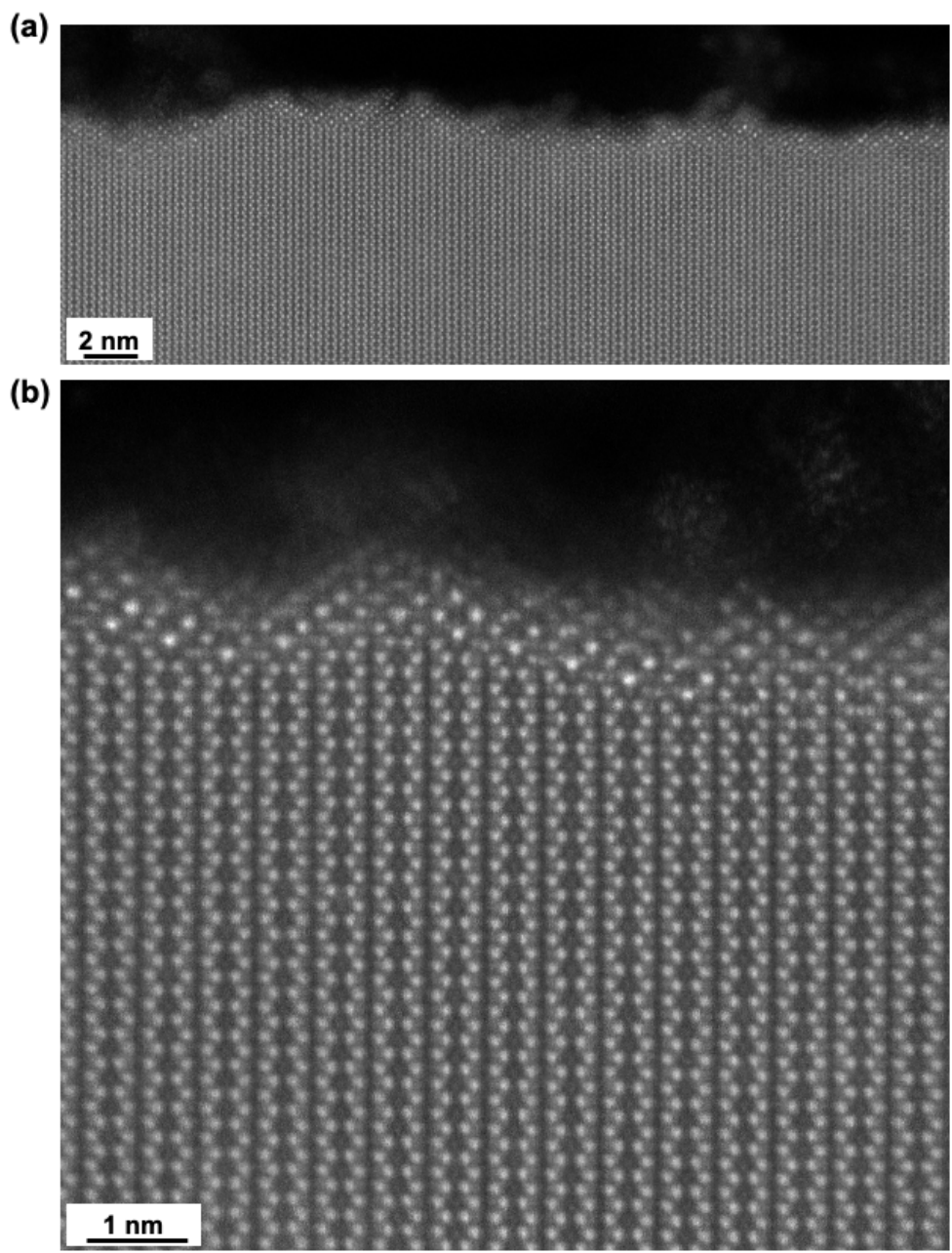

**Figure S2:** HAADF-STEM images of UID Si-doped $\beta$-Ga$_2$O$_3$ (<~$10^{16}$/cm$^3$) grown in 650 ˚C with Si cell temperature at 300 ˚C. (a) Low magnification image showing onset of $\gamma$-phase by the nucleation at bright octahedral Ga sites. (b) Enlarged surface area showing $\gamma$-phase surface layer that is less than 1nm in thickness.

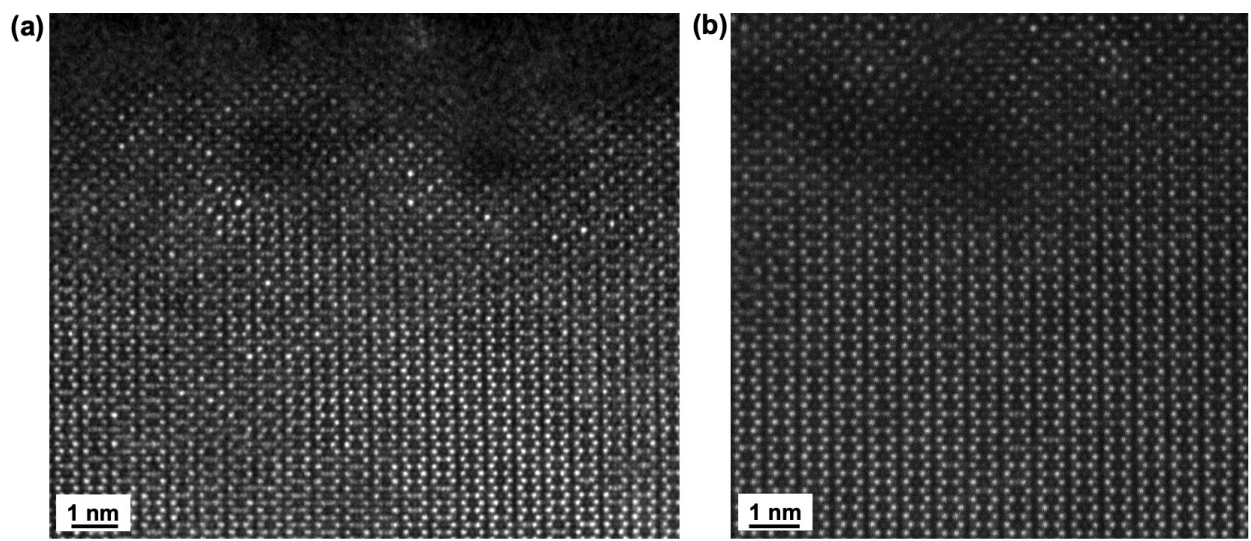

**Figure S3:** HAADF-STEM images of Sn-doped $\beta$-$(Al_{0.15}Ga_{0.85})_2O_3$ (DSn). (a) Image with a slightly larger field of view than shown in Fig. 5(c). Here we can observe the inclusion of $\gamma$-phase on the surface as well as in the film. Higher contrast of octahedral Ga sites and interstitials can also be observed. (b) High density of interstitials is observed in the film.

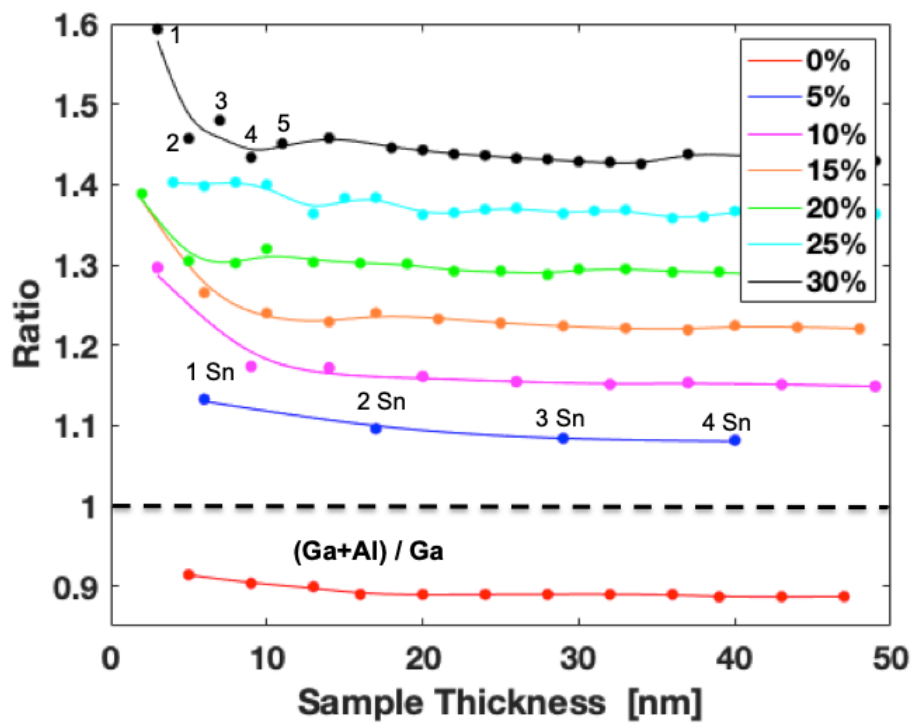

**Figure S4:** The bottom-most plot shows the HAADF contrast ratio of a column having 15% of Al substituted in Ga to a pure Ga column. The reduced contrast is 10%, which will be affected by fluctuations of Al and scanning noise in the image that will reduce the actual contrast to be less than 10%. The six plots above the dotted line shows the contrast ratios of columns that include Sn, Ga and Al to columns that only have Ga and Al, at different Sn concentrations. Note that the first data points for each plot indicate that the contrast is dominated by one single Sn atom. The plot also indicates that thinner samples can benefit further from the enhanced contrast ratio.

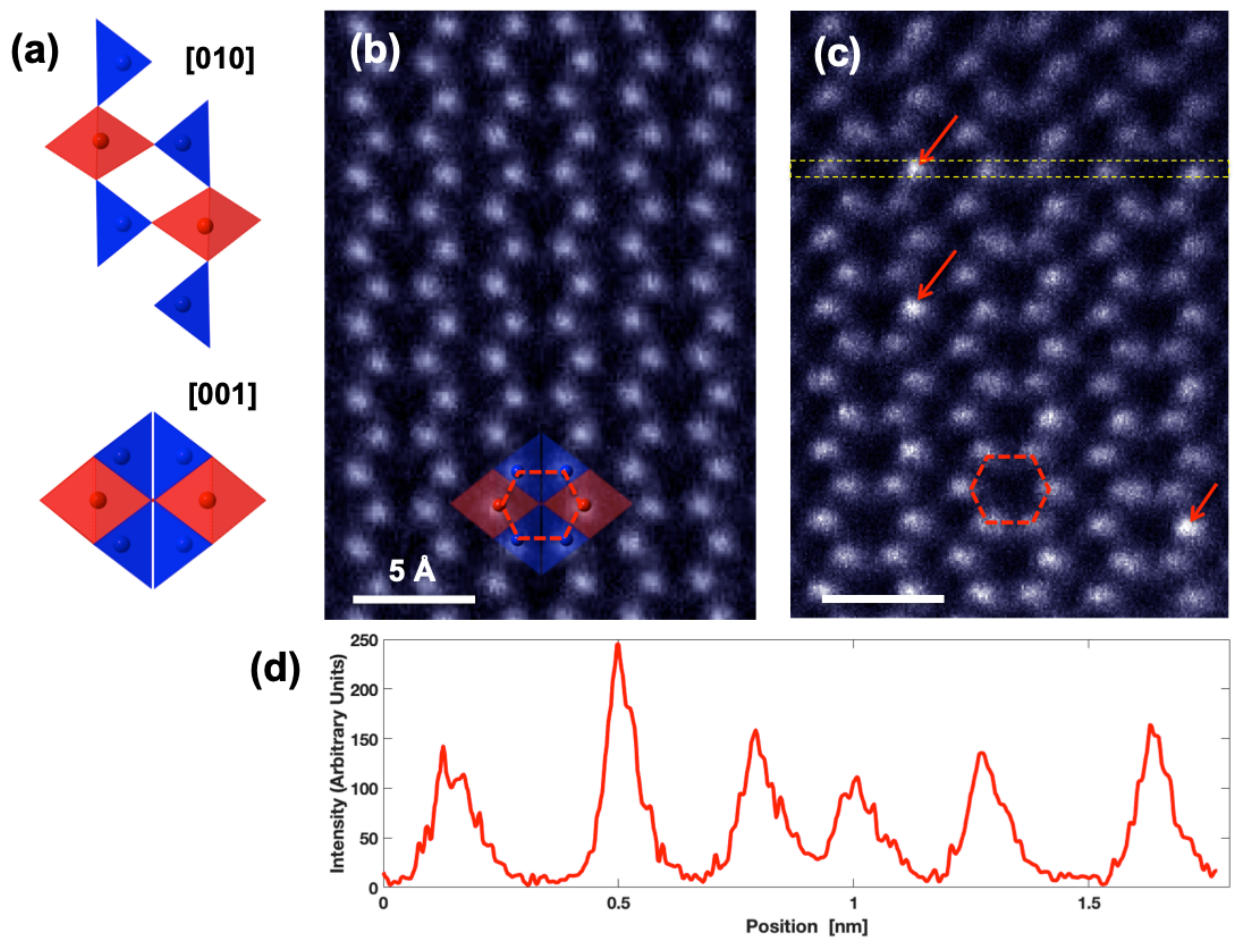

**Figure S5:** (a) There are two sites for Ga: red refers to the octahedral site while blue refers to the tetrahedral sites. In [001] crystal zone we can benefit from seeing projections of purely octahedral sites apart from tetrahedral sites. (b) is a HAADF-STEM image of $\beta$-$(Al_{0.15}Ga_{0.85})_2O_3$ that shows equal contrast of atoms at Ga sites while Sn-doped $\beta$-$(Al_{0.15}Ga_{0.85})_2O_3$ in (c) shows brighter Sn atoms as indicated by red arrows. The structural overlay in (b) corresponds to a projection of Ga atoms on [001] zone axis as shown in (a). This shows that in (c), the Sn-substitution occurs preferentially at the octahedral sites. The intensity integrated over the yellow-dashed area in (c) is plotted as a line profile in (d). The second atomic site from the left shows outstanding intensity compared to the neighboring Ga atomic sites, indicating the substitution of Tin.

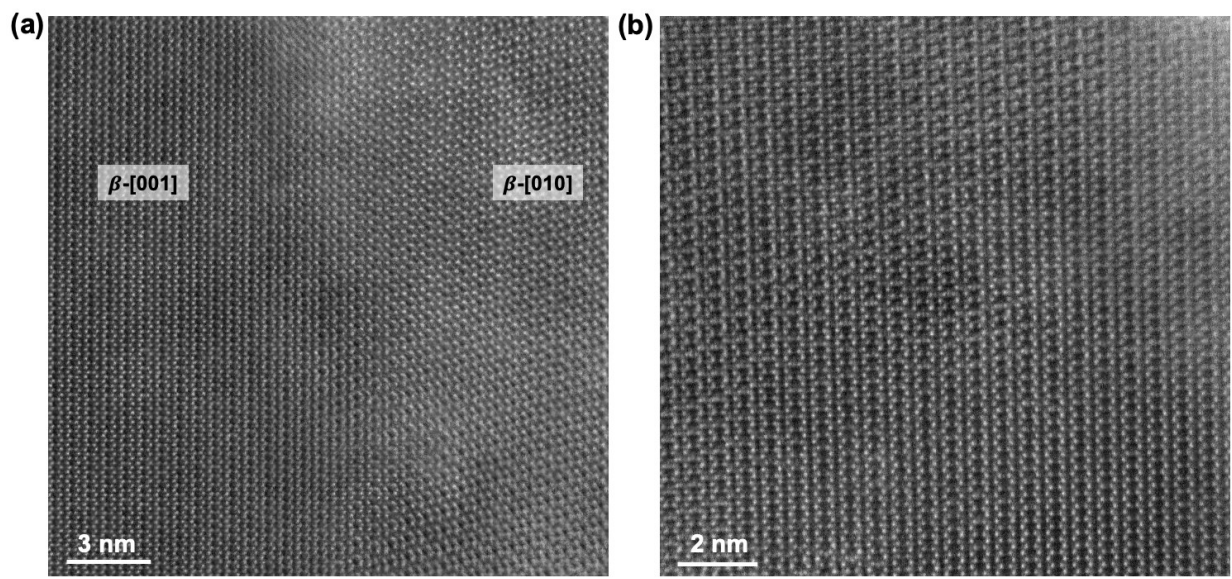

**Figure S6:** HAADF-STEM images of sample A1. (a) Region in the film show coexistence of $\beta$ [001] and [010] phases. (b) Extremely high density of Ga interstitials was observed in several regions of the sample.

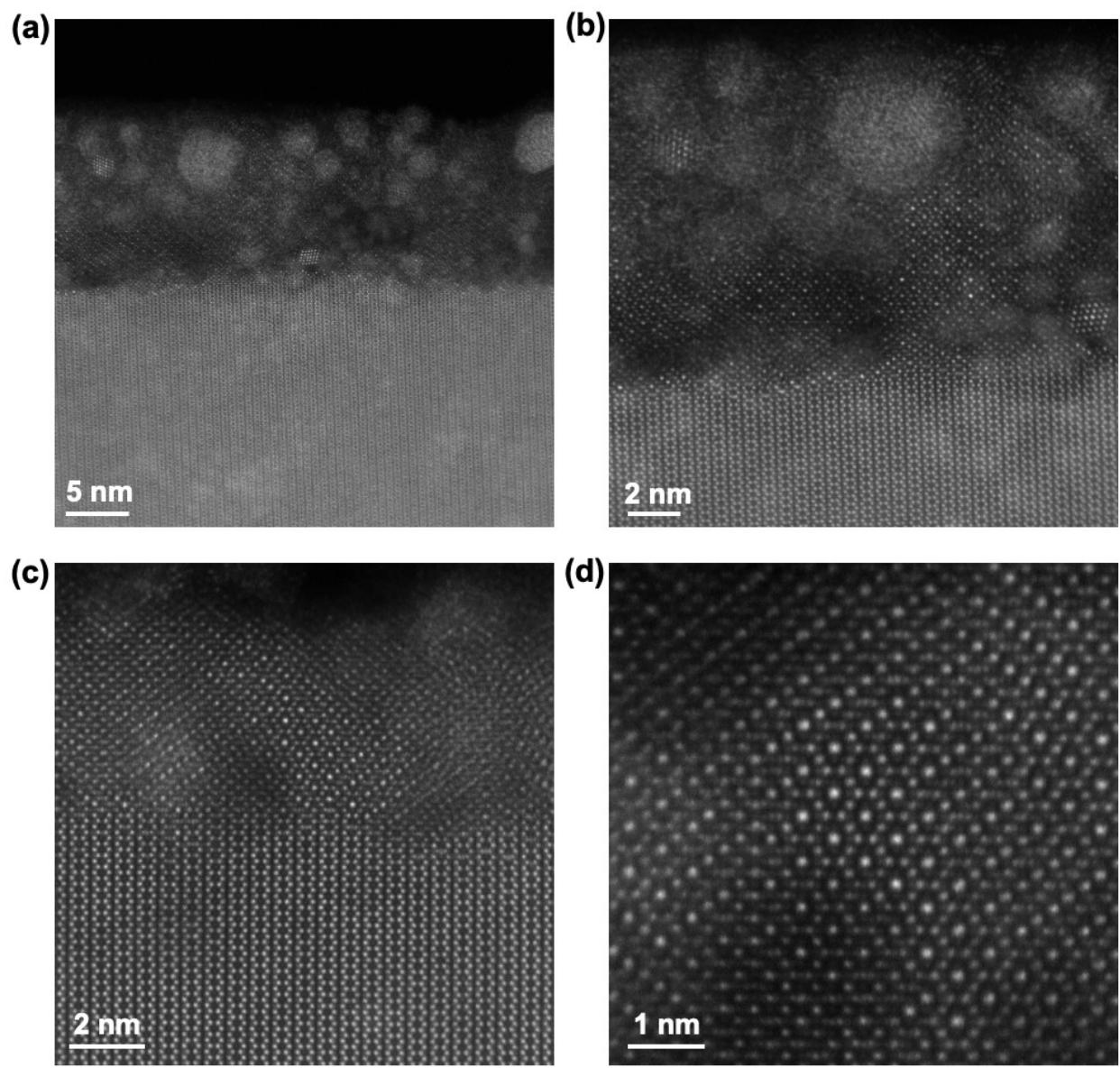

**Figure S7:** HAADF-STEM images of Sn-doped $\beta$-$(Al_{0.15}Ga_{0.85})_2O_3$ (DSn) samples after being heated up to 400 ˚C. (a) and (b) shows newly created $\gamma$-phase on the surface. (c) and (d) shows a region that already had $\gamma$-phase at room temperature. Before heating, it would damage under the electron beam very easily, while after heating we saw no damage. Therefore, we conclude that the $\gamma$-phase became more stable.